\documentclass[traditabstract]{aa}
\usepackage{graphicx}
\usepackage{txfonts}
\usepackage{natbib}
\usepackage[colorlinks=true,citecolor=blue]{hyperref}
\usepackage{xspace}
\usepackage{multirow}
\usepackage{array}

\bibpunct[; ]{(}{)}{;}{a}{}{,}

\newcommand{\as}{\hbox{$^{\prime\prime}$}\xspace}
\newcommand{\lsd}{\hbox{$\lambda/D$}\xspace}

\begin{document}

\title{Calibration of quasi-static aberrations in exoplanet direct-imaging instruments with a Zernike phase-mask sensor\\ II. Concept validation with ZELDA on VLT/SPHERE}
\titlerunning{Zernike phase-mask sensor. II.}

\author{
M. N'Diaye\inst{1} \and
A. Vigan\inst{2,3} \and
K. Dohlen\inst{2} \and 
J.-F. Sauvage\inst{4,2} \and 
A. Caillat\inst{2} \and
A. Costille\inst{2} \and
J. H. V. Girard\inst{3} \and\\
J.-L. Beuzit\inst{5,6} \and 
T. Fusco\inst{4,2} \and
P. Blanchard\inst{2} \and
J. Le Merrer\inst{2} \and
D. Le Mignant\inst{2} \and
F. Madec\inst{2} \and
G. Moreaux\inst{2} \and\\
D. Mouillet\inst{5,6} \and
P. Puget\inst{5,6} \and 
G. Zins\inst{3,5,6}
}

\institute{
Space Telescope Science Institute, 3700 San Martin Drive, Baltimore, MD 21218, USA\\ 
\email{\href{mailto:mamadou@stsci.edu}{mamadou@stsci.edu}} 
\and
Aix Marseille Universit\'e, CNRS, LAM (Laboratoire d'Astrophysique de Marseille) UMR 7326, 13388, Marseille, France 
\and
European Southern Observatory, Alonso de Cordova 3107, Vitacura, Santiago, Chile 
\and 
ONERA, The French Aerospace Lab, BP72, 29 avenue de la Division Leclerc, 92322 Ch\^{a}tillon Cedex, France
\and
CNRS, IPAG (Institut de Plan\'etologie et d'Astrophysique de Grenoble), UMR 5274, B.P. 53, F-38041 Grenoble Cedex 9, France
\and 
Universit\'e Grenoble Alpes, IPAG, 38000 Grenoble, France
}

\date{}
   
\abstract{Warm or massive gas giant planets, brown dwarfs, and debris disks around nearby stars are now routinely observed by dedicated high-contrast imaging instruments that are mounted on large, ground-based observatories. These facilities include extreme adaptive optics (ExAO) and state-of-the-art coronagraphy to achieve unprecedented sensitivities for exoplanet detection and their spectral characterization. However, low spatial frequency differential aberrations between the ExAO sensing path and the science path represent critical limitations for the detection of giant planets with a contrast lower than a few $10^{-6}$ at very small separations (<0.3\as) from their host star. In our previous work, we proposed a wavefront sensor based on Zernike phase-contrast methods to circumvent this problem and measure these quasi-static aberrations at a nanometric level. We present the design, manufacturing, and testing of ZELDA, a prototype that was installed on VLT/SPHERE during its reintegration in Chile. Using the internal light source of the instrument, we first performed measurements in the presence of Zernike or Fourier modes introduced with the deformable mirror. Our experimental results are consistent with the results in simulations, confirming the ability of our sensor to measure small aberrations (<50 nm rms) with nanometric accuracy. Following these results, we corrected the long-lived non-common path aberrations in SPHERE based on ZELDA measurements and estimated a contrast gain of 10 in the coronagraphic image at 0.2\as, reaching the raw contrast limit set by the coronagraph in the instrument. In addition to this encouraging result, the simplicity of the design and its phase reconstruction algorithm makes ZELDA an excellent candidate for the online measurements of quasi-static aberrations during the observations. The implementation of a ZELDA-based sensing path on the current and future facilities (ELTs, future space missions) could facilitate the observation of cold gaseous or massive rocky planets around nearby stars.}


\maketitle
   
\section{Introduction}
\label{sec:introduction}
\defcitealias{N'Diaye2013a}{Paper I}

Circumstellar disks and planetary companions around nearby stars are routinely observed on the ground by several facilities with exoplanet direct-imaging capabilities \citep[e.g.][]{Beuzit2008,Macintosh2008,Guyon2010c,Hinkley2011,Skemer2012,Close2014}. Of these facilities, the instruments VLT/SPHERE and Gemini Planet Imager (GPI) have recently seen first light in 2013-2014, providing unprecedented sensitivity and inner working angle for exoplanet observations \citep{Macintosh2014,vigan2015}. Since their commissioning, they have shed light on known or newly detected planetary companions with insights on their physical characteristics (orbit and mass) and atmospheric chemical features through spectral characterization and photometric and astrometric information \citep{galicher2014b,chilcote2015,vigan2016,maire2016,zurlo2016,bonnefoy2016}. Similar to the recent discovery of 51~Eri~b \citep{Macintosh2015}, large surveys of nearby stars with these instruments are expected to unveil more gas giant planets, providing clues for comparative exoplanetology and enabling a better understanding on the formation and evolution of planetary systems.  

To achieve direct imaging and spectroscopy of companions orbiting nearby stars, these ground-based instruments rely on a combination of extreme adaptive optics (ExAO) system for the fine control of the wavefront errors that are due to atmospheric turbulence and optic imperfections, coronagraphy for starlight suppression, and dedicated observational strategies and post-processing methods to retrieve the signal of the substellar mass companions. With their near-infrared capabilities, these instruments can observe faint planetary-mass companions in thermal emission and study young or massive gaseous planets with contrast ratios down to $10^{-5}-10^{-6}$ at 0.2-0.3\as, corresponding to solar system scales for stars within 100\,pc. 

Differential aberrations between the ExAO sensing path and the science path, so-called non-common path aberrations (NCPA), have been identified as setting high-contrast performance limits for adaptive optics instruments. Their importance was well known \citep[e.g.][]{Fusco2006} at the start of the development of the recently commissioned planet imagers, GPI and SPHERE, and various strategies were implemented to minimize them. In particular, SPHERE has a differential tip-tilt correction system, using a camera close to the coronagraph as an image position sensor, to ensure that any differential image movement is compensated for. Thermally induced differential defocus was minimized by design, and on-coronagraph focus is optimized at the start of observations. Higher-order aberrations, responsible for the residual quasi-static speckles that limit the high-contrast performance of the instrument, were first of all minimized by tightly specifying all optics in the differential path, and second by implementing AO calibration strategies to minimize residual aberrations. For SPHERE the differential optics worked better than specified, and consequently, the adopted calibration strategy, which is based on phase diversity techniques \citep{sauvage2007}, was not found to improve the final image quality and was finally discarded. Still, the remaining NCPA are on the order of a few tens of nanometers, preventing coronagraphs from achieving their ultimate performance. These wavefront errors can be split into two contributions: the long-timescale aberrations that are due to the optical surface errors or misalignments in the instrument optical train and the slowly varying instrumental aberrations that are caused by thermal or opto-mechanical deformations  as well as moving optics such as atmospheric dispersion correctors \citep[e.g.][]{macintosh2005,martinez2012,Martinez2013}. They lead to static and quasi-static speckles in the coronagraphic images, which represent critical limitations for the detection and observation of older or lighter gaseous planets at smaller separations. More precise measurement strategies are required to measure and correct for these small errors with accuracy and achieve deeper contrast (down to $10^{-7}$, representing the ultimate contrast limit of these instruments) for the observation of the faintest companions.
Other unforeseen limiting effects have also been experienced with these new instruments of unprecedented performance. In particular, the low-wind effect (LWE) where piston patterns appear across the spiders in monolithic-pupil telescopes, has turned out to be devastating as it systematically occurs when the outdoor wind speed (i.e., 30\,m below the VLT platform) is below 1 to 3\,m/s \citep{Sauvage2016}. Around 20\% of observing time is affected by this effect, causing strong speckle in the first Airy ring and even splitting of the main peak into up to four peaks of varying intensity. The cause of this effect has been identified as being energetic transfer between spiders that are radiatively cooled by the night sky, and the slowly passing dome air. While strategies such as repainting spiders using a low-emissivity paint are being studied to control the root cause of this particular case of dome seeing, real-time phase measurements may be needed to correct for its effect.

Over the past few years, several methods have been proposed to circumvent the NCPA problem \citep[e.g.][]{gonsalves1982,wallace2010,Paul2013}. We have proposed the use of a Zernike phase mask sensor to calibrate the NCPA seen by the coronagraph in exoplanet direct imagers \citep[][hereafter \citetalias{N'Diaye2013a}]{N'Diaye2013a}. This phase-contrast method uses a phase-shift mask to modulate the phase differential aberrations into intensity variations in the pupil plane. Since differential aberrations in exoplanet imagers are small, a linear or quadratic relation between the wavefront errors and the pupil intensity enables reconstructing the differential aberrations at nanometric accuracy with a simple, fast algorithm, making calibration in real time possible.

Zernike sensors have been explored in astronomy to address various instrumentation aspects, such as wavefront sensing in adaptive optics systems or cophasing of telescope segmented primary mirror \citep{Bloemhof2003,Bloemhof2004a,Dohlen2004,Surdej2010,2011SPIE.8126E..11W,Vigan2011}. Recently, the Zernike sensor has been adopted for the WFIRST mission to measure low-order aberrations in its coronagraphic instrument and control pointing errors and focus drifts on the coronagraphic mask \citep{Spergel2013a,Spergel2015,Zhao2014}. Laboratory demonstration of the concept have been carried out in this context \citep{Shi2015}. We have also performed preliminary tests and obtained encouraging results of the Zernike sensor on the coronagraphic testbed in Marseille \citep{N'Diaye2012b,N'Diaye2014b,Dohlen2013}. However, and to the best of our knowledge, no experimental validation has been performed on a real instrument for the measurement of coronagraphic aberrations. 

In this paper, we propose to validate the Zernike phase-mask sensor for the measurement of differential aberrations in a real exoplanet direct-imaging instrument, SPHERE at the VLT. We designed a sensor called ZELDA (which stands for Zernike sensor for Extremely Low-level Differential Aberrations) and installed a manufactured prototype in SPHERE during its reintegration at the ESO Paranal observatory in 2014. We present the first experimental results, using the internal light source of the instrument. After recalling the principle and formalism of the Zernike sensor, we present the design, properties, and manufacturing of ZELDA. We then evaluate the performance of our wavefront sensor in the presence of different aberration modes and analyze its sensitivity to spectral bandpass. We finally present the first results of ZELDA-based wavefront correction on VLT/SPHERE for the long-lived speckles and derive the contrast gain achieved with this calibration.   

\section{ZELDA sensor}\label{sec:ZELDA_sensor}
\subsection{Principle and formalism}

The ZELDA sensor is based on phase-contrast techniques that were proposed by \citet{Zernike1934} to measure NCPA in high-contrast imaging instruments with nanometric accuracy. We briefly recall the principle and formalism of this Zernike sensor that were detailed in \citetalias{N'Diaye2013a}.

\begin{figure}
    \centering
    \includegraphics[width=0.5\textwidth]{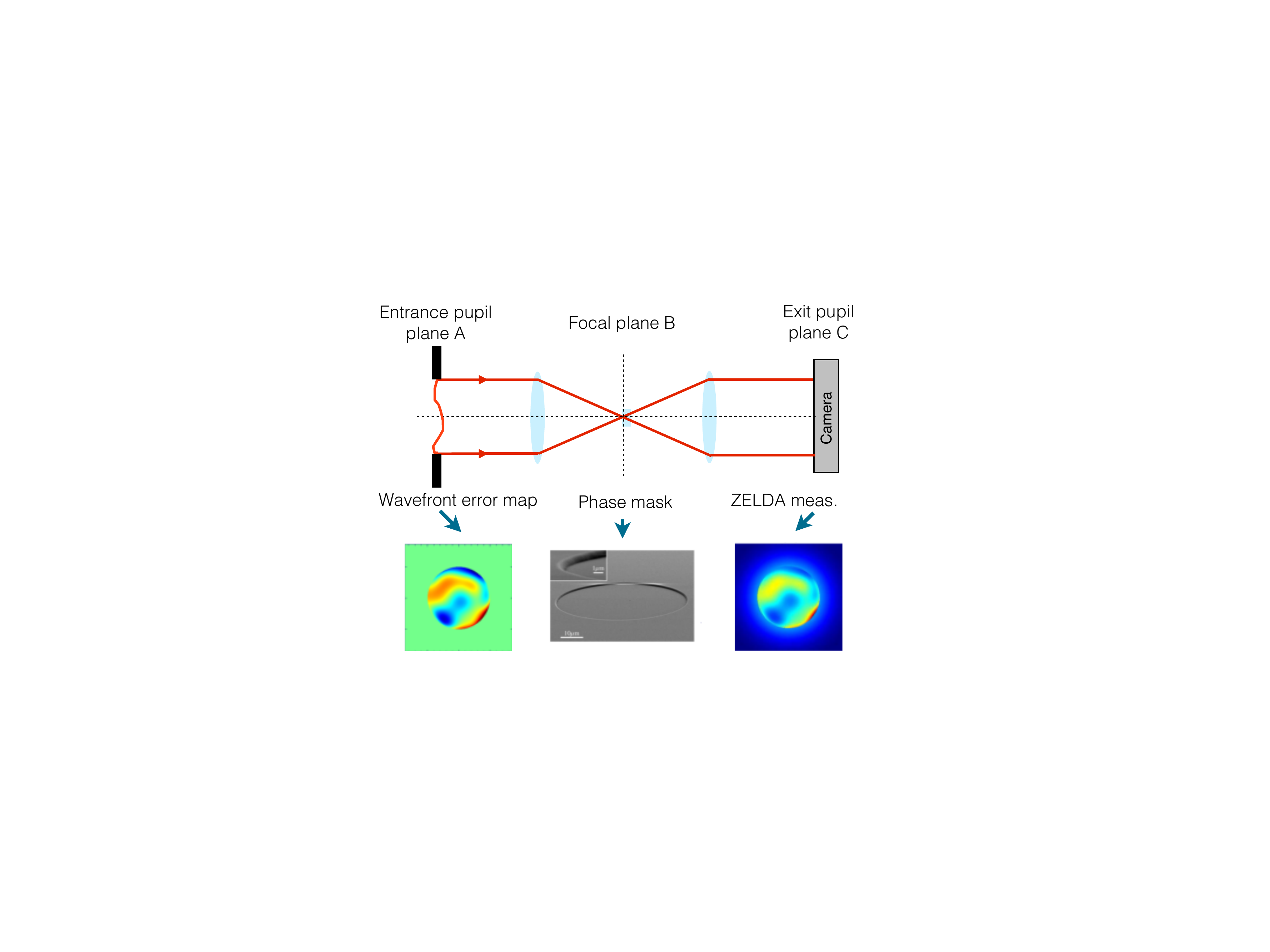}
    \caption{Principle of the ZELDA analysis with wavefront errors in the entrance pupil plane A to estimate, a phase mask centered on the stellar signal in the following focal plane B, and the intensity measurement in the re-imaged pupil plane C. A linear or quadratic reconstruction of the aberrations is performed from the recorded intensity with nanometric accuracy.}
    \label{fig:zelda_scheme}
\end{figure}

This sensor uses a focal plane phase mask to produce interference between a reference wave created by the mask and the phase errors present in the system (Fig. \ref{fig:zelda_scheme}). As a result, this sensor converts the aberrations in the entrance pupil into intensity variations in the exit pupil. This phase-to-intensity conversion depends on the mask characteristics, that is, the diameter $d$ and the depth that is related to the sensor phase delay $\theta$. In the following, $\lambda$ and $D$ denote the wavelength of observation and the telescope aperture diameter. We recall the expression of the ZELDA signal $I_C$ as a function of the phase error $\varphi$ for a given pixel in the entrance pupil:
\begin{equation}
I_C = P^2 + 2b^2(1-\cos\theta) + 2Pb\left [\sin\varphi\sin\theta - \cos\varphi(1-\cos\theta) \right ]\,,
\label{eq:intensity_C}
\end{equation}
where $P$ and $b$ denote the amplitude pupil function and the amplitude diffracted by the focal plane phase mask of diameter $d$. For small phase errors, we can consider a quadratic case where we only maintain first- and second-order terms of $\varphi$ in the Taylor expansion term. Then we have
\begin{equation}
I_C = P^2 +2b^2(1 - \cos\theta) + 2Pb \left [\varphi \sin\theta - (1 - \varphi^2/2)(1 - \cos\theta) \right ]\,.
\label{eq:intensity_C_small_wfe}
\end{equation}
The phase can be recovered from the intensity by solving this second-order equation. For a phase mask with depth $\theta=\pi/2$ and angular diameter 1.06\,$\lambda_0/D$, where $\lambda_0$ denotes the wavelength of design, $b$ ranges between 0.4 and 0.6 (see Fig. 3 of \citetalias{N'Diaye2013a}). Assuming a normalized amplitude in the entrance pupil $P=1$, we obtain the solution
\begin{equation}
\varphi=-1 + \sqrt{3 - 2b - (1 - I_C)/b}\,.
\label{eq:phase_classical_mask}
\end{equation}
The formalism of the Zernike phase-mask sensor is valid for any aperture geometry, including primary mirror segmentation, central obstruction, or spider struts, since the intensity measurement and the phase reconstruction is performed inside the geometric pupil at any given point. This property makes our sensor particularly interesting for measurements in the context of a deformable mirror (DM) with dead actuators as in SPHERE, since their respective points do not alter the measurements of the other points inside the geometric pupil.

\subsection{Design and properties}
\label{subsec:design}

\begin{figure}
    \centering
    \includegraphics[width=0.5\textwidth]{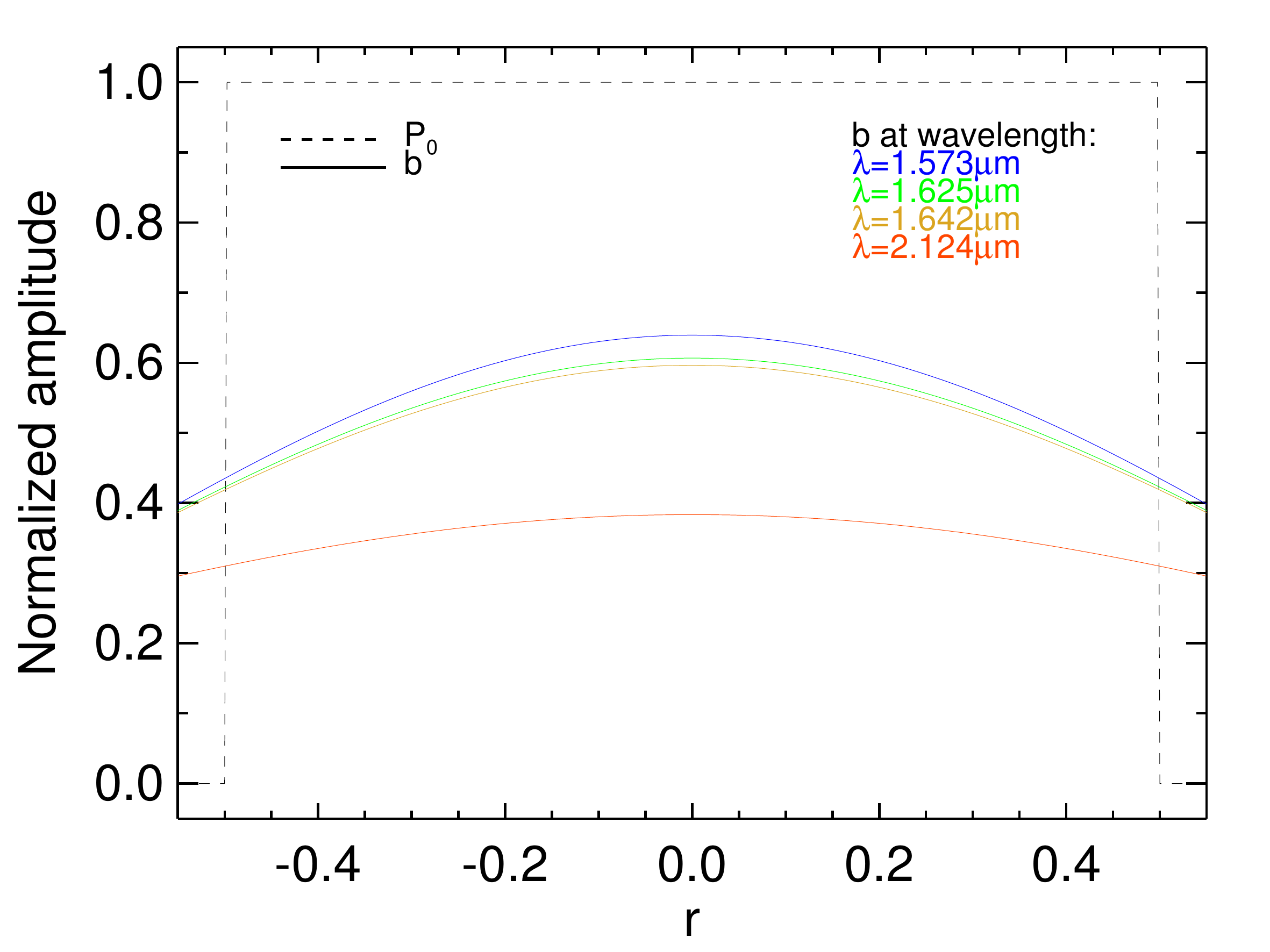}
    \caption{Radial profile of the amplitude $b$ diffracted by a mask of size 1.087\,$\lambda_0/D$ and phase shift $\theta=0.444\pi\lambda_0/\lambda$ at the central wavelength of the filters that are used during our tests with SPHERE. The dashed line defines the entrance pupil function $P_0$.}
    \label{fig:zelda_b-profile}
\end{figure}

Our design for VLT/SPHERE corresponds to the case where $d=1.076\,\lambda_0/D$ and $\theta=0.440\pi\lambda_0/\lambda$ with $\lambda_0=1.642\,\mu$m. With such a mask diameter, $b$ has a chromatically dependant profile similar to that of an Airy pattern twice the size of the pupil (Fig. \ref{fig:zelda_b-profile}). As an illustration, we assume a normalized entrance pupil plane amplitude and measurements are performed at $\lambda=\lambda_0$ and then, the phase error at any given point in the pupil is then reduced to
\begin{equation}
\varphi=-1.208+1.230 \sqrt{2.590 -1.626b - 0.813(1-I_C)/b}\,.
\label{eq:phase}
\end{equation}

\begin{figure}
    \centering
    \includegraphics[width=0.5\textwidth]{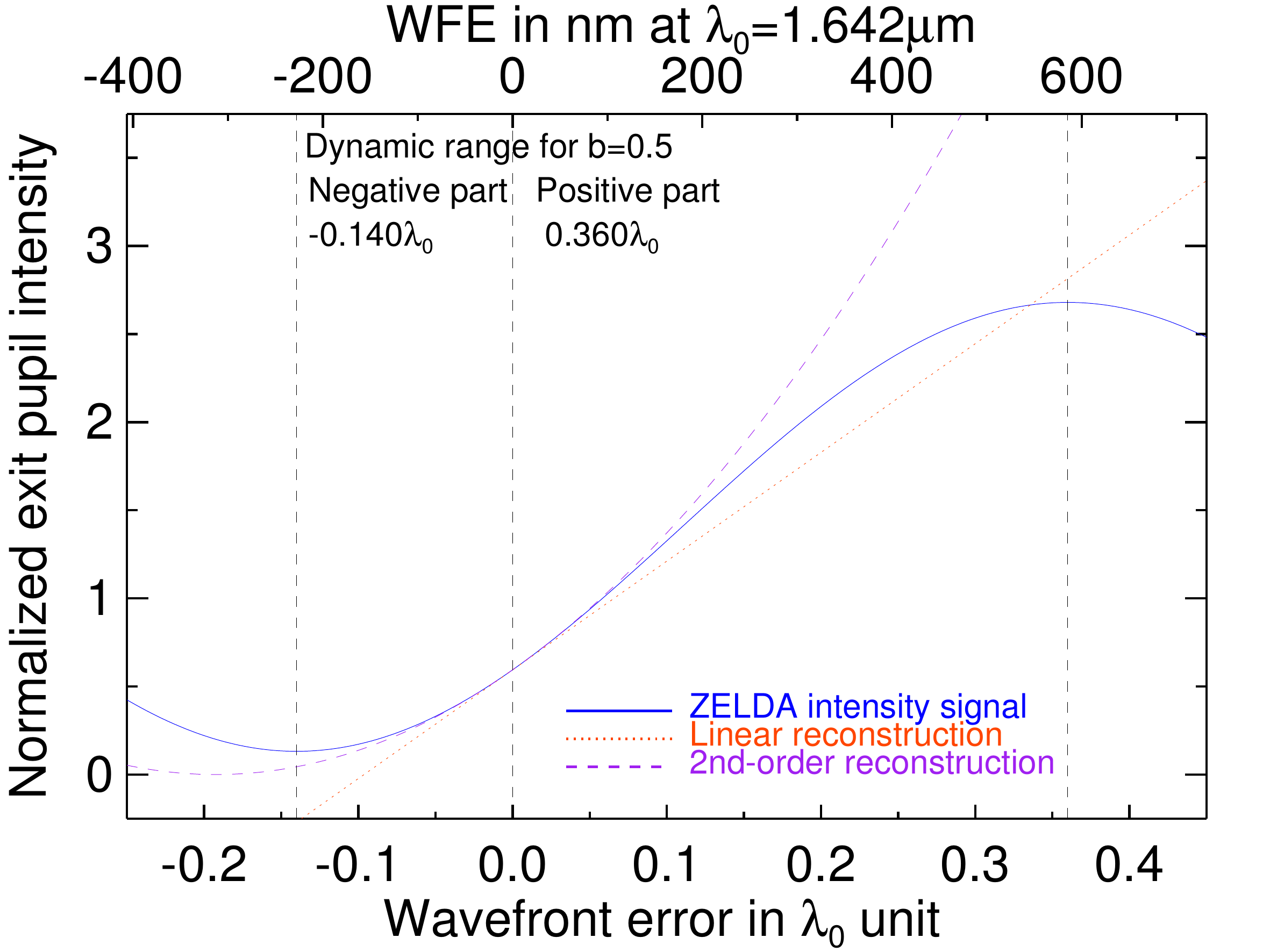}
    \caption{ZELDA pupil plane intensity as a function of phase aberration for a given pixel in the pupil, assuming a mask-diffracted wave amplitude $b=0.5$. The dynamic range of our sensor is represented with dashed vertical lines. The linear and second-order phase reconstruction are displayed with dot and dashed lines.}
    \label{fig:zelda_dynamic_range}
\end{figure}

Figure \ref{fig:zelda_dynamic_range} shows the ZELDA signal at $\lambda=\lambda_0$ as a function of the wavefront error for a given pixel with $b=0.5$. The intensity received by a pixel depends on the wavefront error (WFE) location of that pixel on a sinusoidal function. However, the sinusoid is not symmetric about zero aberration, giving rise to an asymmetric dynamic range defined by the monotonic range around zero. The limits of the dynamic range are given by the changes of gradient sign of $I_C$, that is, $dI_C/d\varphi=0$. In our mask design, the dynamic of the sensor ranges between -0.14\,$\lambda_0$ and 0.36\,$\lambda_0$, as illustrated with the vertical lines in Fig. \ref{fig:zelda_dynamic_range}. 

\subsection{Prototype}
\label{subsec:prototype}
The phase mask consists of a circular shape machined into the front face of a fused silica substrate by the aid of photolithographic reactive ion etching. This subtractive process, which has been experimented with and optimized in the context of the Roddier \& Roddier coronagraph \citep{N'Diaye2010,N'Diaye2011} and Zernike wavefront sensors \citep{Dohlen2006,N'Diaye2014b}, has been found superior to the more classical additive process where SiO$_2$ is deposited onto a fused silica substrate \citep{Guyon1999}. While the ion-etching process offers extremely steep edges and precisely defined phase steps, it is also monolithic, avoiding any interfaces between materials that might give rise to spurious interference effects.

\begin{figure}
    \centering
    \includegraphics[width=0.35\textwidth]{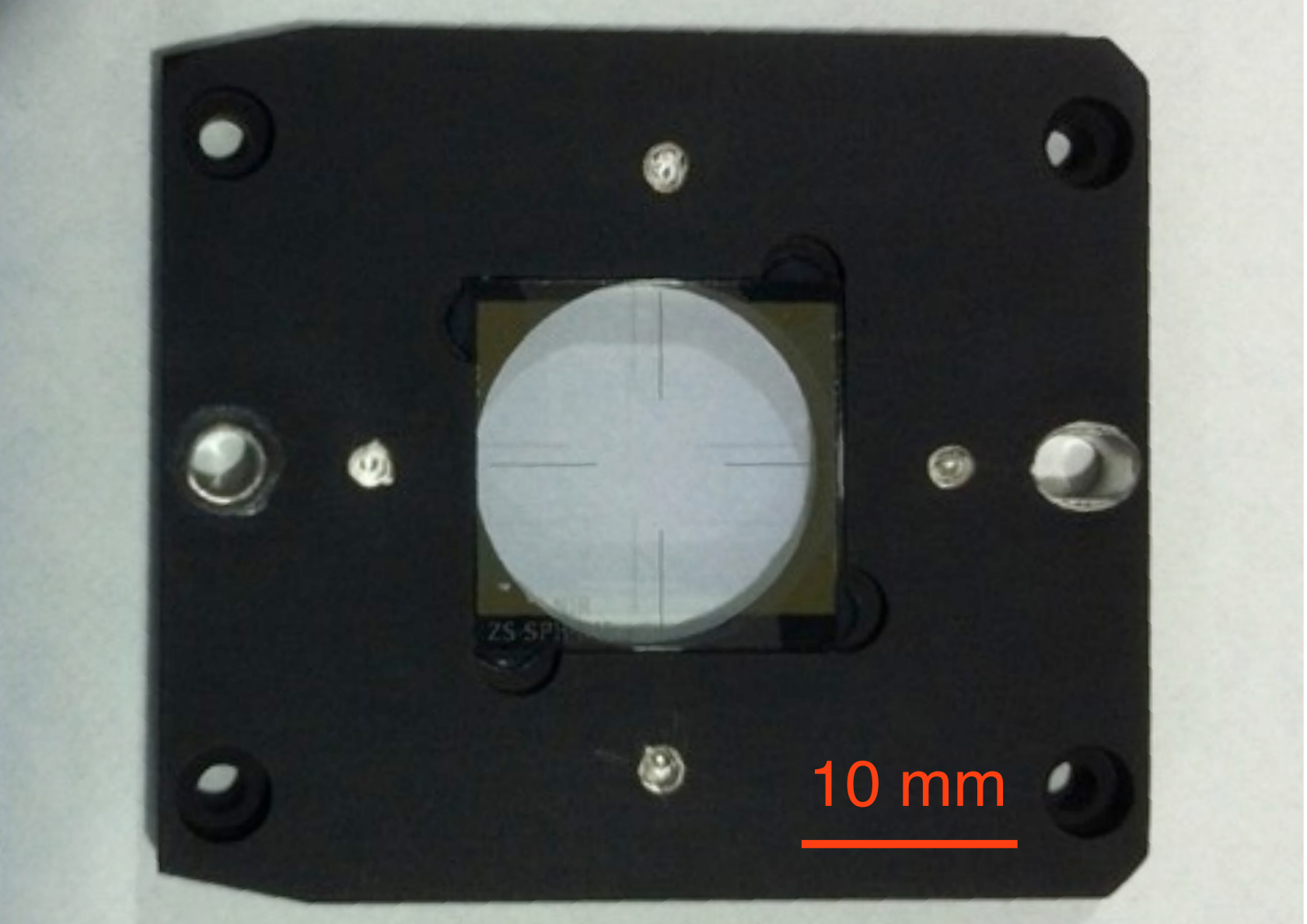}
    \caption{Picture of the ZELDA prototype in its assembly before installation in the SPHERE coronagraph wheel in April 2014.}
    \label{fig:zelda_picture}
\end{figure}

The phase mask was manufactured by the SILIOS company in two steps. First, a circular hole was generated and transferred into photoresist by UV photolithographic projection, leaving the surface to be machined naked. Then, reactive ion etching was applied by exposing the surface to SF$_6$ gas. A picture of our prototype is shown in Fig. \ref{fig:zelda_picture}.

\begin{figure}
    \centering
    \includegraphics[width=0.5\textwidth]{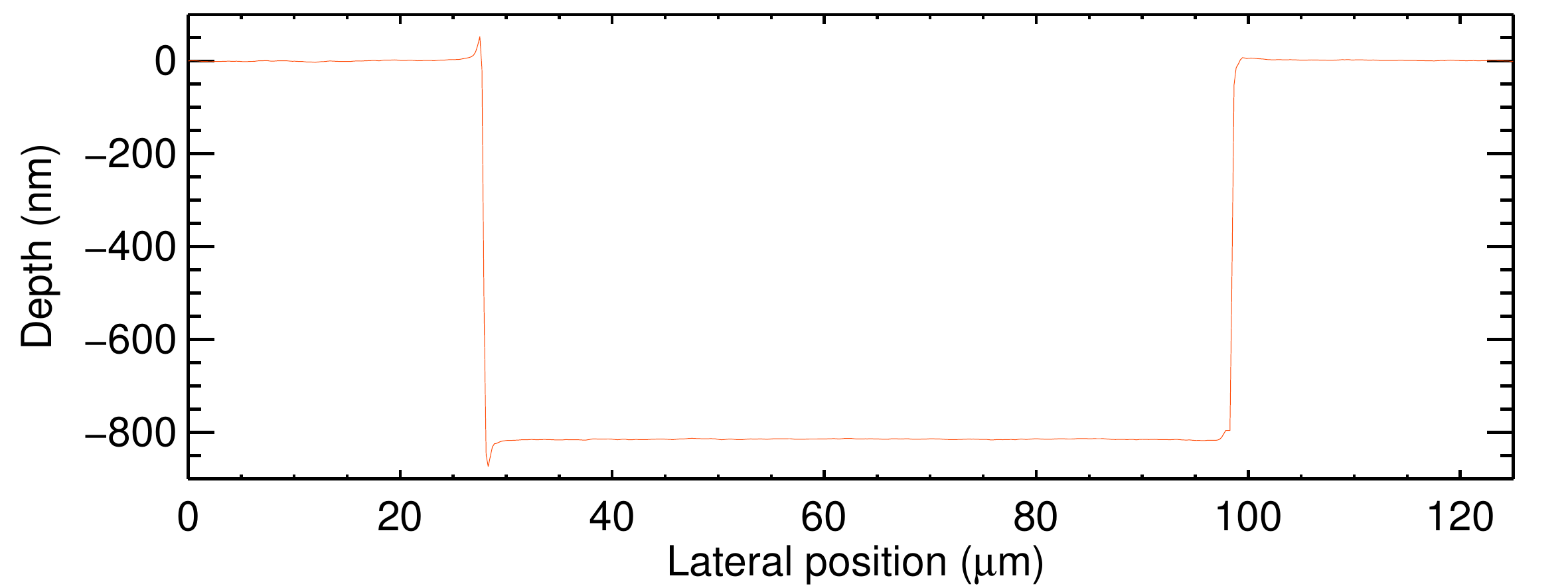}
    \includegraphics[width=0.5\textwidth]{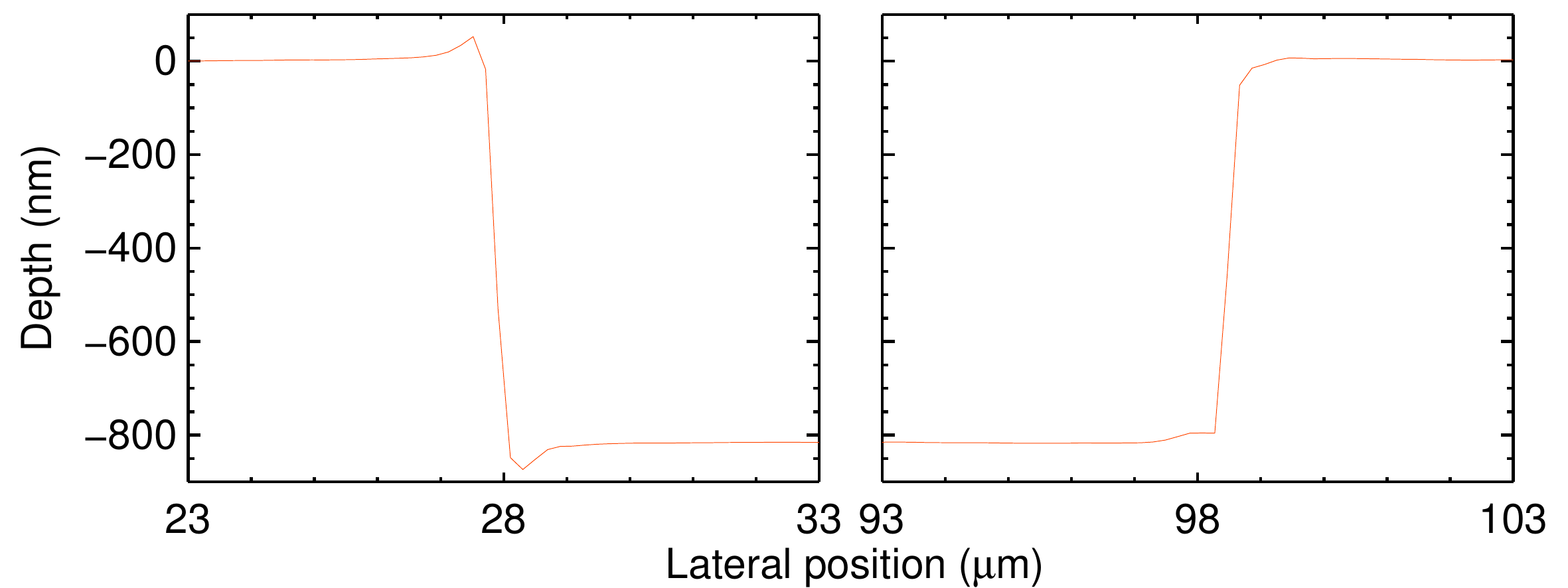}
    \includegraphics[width=0.5\textwidth]{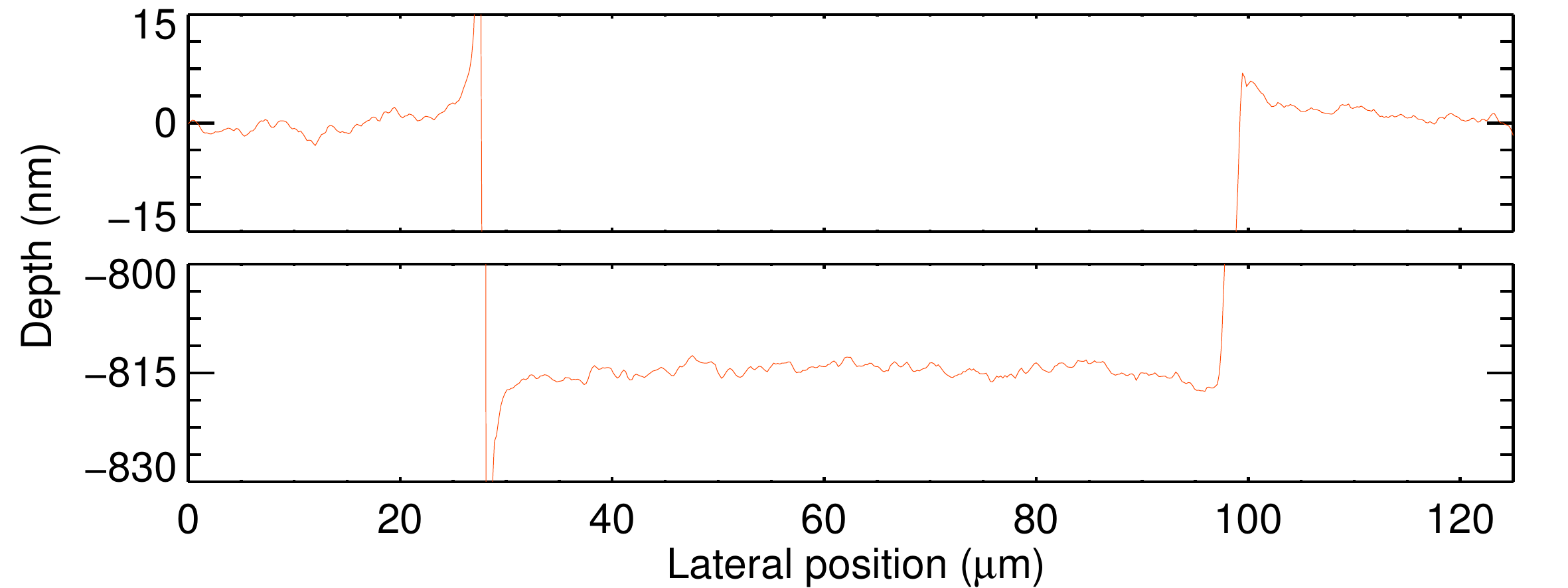}
    \caption{Profile of the phase mask measured by profilometry. Top: general profile. Middle: horizontal zoom at the mask transition areas. Bottom: vertical zoom on the roughness of the different mask steps.}
    \label{Fig_Phase_Mask_Profile}%
\end{figure}

The excellent shape of masks made by this procedure is evident from the optical profilometer made using a Wyko interference microscope and presented in Fig.\,\ref{Fig_Phase_Mask_Profile}. Table\,\ref{table:phase_mask} gives the measured depth and diameter for this prototype, showing good agreement with our specifications: a relative error better than 1\% is achieved for both dimensions. The root mean square (rms) roughness within the machined area is 0.9\,nm, which proves identical to that of the substrate outside of the mask. While a slight rounding of the edges can be seen, the transition zone is less than 1\,$\mu$m wide, corresponding to about 1\% of the mask diameter, which is within the specifications and the accuracy range of the manufacturing process. The slight offset of the mask shape does not impact the ZELDA performance as the wavefront reconstruction algorithm is adjusted with the characteristics of the manufactured phase mask to achieve optimal phase error measurements. 

\begin{table}
\caption{Dimensions of the phase mask from specifications and laboratory measurements using a Wyko interferometer, and relative differences in percent between both values.} 
\centering
\begin{tabular}{c c c c}
\hline\hline
\multirow{2}{*}{Parameters} & \multirow{2}{*}{Specification} & Laboratory & Relative\\
 &  & measurement & difference\\
\hline
diameter $d$ & 69.9\,$\mu$m & 70.7\,$\mu$m & 1.1\,\% \\
depth $z$    & 814.1\,nm    & 814.6\,nm    & 0.1\,\% \\
\hline
\end{tabular}
\label{table:phase_mask}
\end{table}

\section{ZELDA performance}
\label{sec:zelda_performance}

In 2015 we performed multiple validation tests of the ZELDA wavefront sensor in SPHERE. This Zernike sensor was installed in the infrared coronagraphic wheel of the instrument during its reintegration in Paranal in 2014. For our test purposes, the coronagraphic mask is replaced with ZELDA and the system is set up in pupil-imaging mode to perform our phase aberration measurements. During these tests, we introduced Zernike and Fourier modes on the high-order DM of SPHERE, and we compared the ZELDA measurements with the theoretical modes to estimate the performance. The modes were chosen to cover low-order aberrations and spatial frequencies, which are the main components of the NCPA that we wish to correct to improve the performance of an ExAO coronagraphic system. They were introduced at various amplitudes to explore the dynamic range of the sensor. In this section we describe the data acquisition and analysis (Sect.~\ref{sec:data_analysis_processing}) before reviewing the results obtained with different modes (Sect.~\ref{sec:measurements_comparison_theory}) and the sensitivity to spectral bandpass (Sect.~\ref{sec:sensitivity_spectral_bandpass}).

\subsection{Data acquisition and processing}
\label{sec:data_analysis_processing}

\subsubsection{Calibration of the aberrations: sensitivity factor}
\label{sec:calib_aberrations}

For optimal performance, the modes were introduced in the system in closed-loop mode using an offset to the reference slopes of the Shack-Hartmann (SH) wavefront sensor (WFS) of SPHERE. This approach allowed us to benefit from an extremely high-quality point-spread function (Strehl ratio > 90\% at $\lambda_0$) at the level of the ZELDA phase mask, since the system works fully in closed-loop mode. We used the inverse of the influence matrix of the system to convert the phase that we wished to introduce at the surface of the DM into a voltage offset at the level of the individual actuators. This voltage vector was then converted into reference slope offsets using the interaction matrix of the system, which is calibrated daily as part of the calibration plan of the instrument. 

To be accurate, this procedure requires a calibration step to precisely estimate the amount of aberrations that are introduced for a given offset on the reference slopes, which from now on we refer to as the sensitivity factor. In other words, we wish to determine the real amount of aberrations that are introduced when we request 1~nm~RMS of aberrations. This amplitude calibration was performed by introducing a known aberration of a given amplitude onto the DM and measuring the loss of Strehl ratio induced at the level of the PSF in the focal plane. For small amounts of aberrations, Mar{\'e}chal's approximation tells that the Strehl ratio can be expressed as $S_{r} = e^{-\sigma_{\phi}^2}$, with $\sigma_{\phi}^2$ the variance of the wavefront error. In practice, we introduced a ramp of focus instead of a single value to make the measurement more robust to noise. Then an inverse exponential function with three free parameters was fit to the Strehl values measured for each focus value. The free parameters correspond to the position of the maximum, the sensitivity factor that we searched for, and a scaling factor that is necessary to take the static aberrations of the system into account. 

When performing this calibration, the AO system is operated in particular conditions. First, the light source is an internal calibration source, hence allowing an operation with very high flux that would correspond to a star of negative $R$-band magnitude. All potential problems related to noise propagation are therefore completely avoided. Second, the aberrations to be compensated for are considered static, or with characteristic evolution time so small with respect to the AO-loop frame-rate (1.2 kHz) that any temporal filtering by the loop can be considered as negligible. To project the aberration phase map onto the reference slopes, the calibrated interaction matrix of the AO loop is used. This matrix represents the sensitivity of the SH wavefront sensor to all the modes controlled by the system, it therefore accounts for the system response as accurately as possible. As a consequence of this procedure, the sensitivity factor is valid for any other mode, although it has been calibrated only on the focus.

\subsubsection{Data acquisition in SPHERE}
\label{sec:data_acquisition_sphere}

\begin{table*}
  \caption{Summary of data acquired}
  \label{tab:data_acquisition_log}
  \centering
  \begin{tabular}{lcccccc}
  \hline\hline
  Mode                     & Date       & Amplitude range  & Steps & Filter   & Wavelength & Bandwidth \\
                           &            & (nm PtV)         &       &          & (nm)       & (nm)      \\
  \hline
  \multicolumn{7}{c}{Zernike modes} \\
  \hline
  Tip, Z1                  & 2015-11-30 & -250 \ldots +600 & 35 & Fe~{\sc ii} & 1642       & 24        \\
  Tilt, Z2                 & 2015-11-30 & -250 \ldots +600 & 35 & Fe~{\sc ii} & 1642       & 24        \\
  Focus, Z3                & 2015-11-30 & -250 \ldots +600 & 35 & Fe~{\sc ii} & 1642       & 24        \\
  Astig x, Z4              & 2015-11-30 & -250 \ldots +600 & 35 & Fe~{\sc ii} & 1642       & 24        \\
  Astig y, Z5              & 2015-11-30 & -250 \ldots +600 & 35 & Fe~{\sc ii} & 1642       & 24        \\
  Coma x, Z6               & 2015-12-16 & -250 \ldots +600 & 35 & Fe~{\sc ii} & 1642       & 24        \\
  Coma y, Z7               & 2015-12-16 & -250 \ldots +600 & 35 & Fe~{\sc ii} & 1642       & 24        \\
  Trefoil x, Z8            & 2015-12-17 & -250 \ldots +600 & 35 & Fe~{\sc ii} & 1642       & 24        \\
  Trefoil y, Z9            & 2015-12-17 & -250 \ldots +600 & 35 & Fe~{\sc ii} & 1642       & 24        \\
  Spherical, Z10           & 2015-12-17 & -250 \ldots +600 & 35 & Fe~{\sc ii} & 1642       & 24        \\
  \hline
  \multicolumn{7}{c}{Fourier modes} \\
  \hline
  Fourier, x, 1 cyc./pup.  & 2015-12-01 & -250 \ldots +600 & 35 & Fe~{\sc ii} & 1642       & 24        \\
  Fourier, x, 2 cyc./pup.  & 2015-12-01 & -250 \ldots +600 & 35 & Fe~{\sc ii} & 1642       & 24        \\
  Fourier, x, 3 cyc./pup.  & 2015-12-01 & -250 \ldots +600 & 35 & Fe~{\sc ii} & 1642       & 24        \\
  Fourier, x, 4 cyc./pup.  & 2015-12-01 & -250 \ldots +600 & 35 & Fe~{\sc ii} & 1642       & 24        \\
  Fourier, x, 5 cyc./pup.  & 2015-12-01 & -250 \ldots +600 & 35 & Fe~{\sc ii} & 1642       & 24        \\
  Fourier, x, 6 cyc./pup.  & 2015-12-01 & -250 \ldots +600 & 35 & Fe~{\sc ii} & 1642       & 24        \\
  Fourier, x, 7 cyc./pup.  & 2015-12-01 & -250 \ldots +600 & 35 & Fe~{\sc ii} & 1642       & 24        \\
  Fourier, x, 8 cyc./pup.  & 2015-12-01 & -250 \ldots +600 & 35 & Fe~{\sc ii} & 1642       & 24        \\
  Fourier, x, 9 cyc./pup.  & 2015-12-01 & -200 \ldots +400 & 31 & Fe~{\sc ii} & 1642       & 24        \\
  Fourier, x, 10 cyc./pup. & 2015-12-01 & -135 \ldots +135 & 21 & Fe~{\sc ii} & 1642       & 24        \\
  \hline
  Fourier, y, 1 cyc./pup.  & 2015-12-17 & -250 \ldots +600 & 35 & Fe~{\sc ii} & 1642       & 24        \\
  Fourier, y, 2 cyc./pup.  & 2015-12-17 & -250 \ldots +600 & 35 & Fe~{\sc ii} & 1642       & 24        \\
  Fourier, y, 3 cyc./pup.  & 2015-12-17 & -250 \ldots +600 & 35 & Fe~{\sc ii} & 1642       & 24        \\
  Fourier, y, 4 cyc./pup.  & 2015-12-18 & -250 \ldots +600 & 35 & Fe~{\sc ii} & 1642       & 24        \\
  Fourier, y, 5 cyc./pup.  & 2015-12-18 & -250 \ldots +600 & 35 & Fe~{\sc ii} & 1642       & 24        \\
  Fourier, y, 6 cyc./pup.  & 2015-12-18 & -250 \ldots +600 & 35 & Fe~{\sc ii} & 1642       & 24        \\
  Fourier, y, 7 cyc./pup.  & 2015-12-19 & -250 \ldots +600 & 35 & Fe~{\sc ii} & 1642       & 24        \\
  Fourier, y, 8 cyc./pup.  & 2015-12-19 & -250 \ldots +600 & 35 & Fe~{\sc ii} & 1642       & 24        \\
  Fourier, y, 9 cyc./pup.  & 2015-12-19 & -200 \ldots +200 & 27 & Fe~{\sc ii} & 1642       & 24        \\
  Fourier, y, 10 cyc./pup. & 2015-12-19 & -135 \ldots +135 & 21 & Fe~{\sc ii} & 1642       & 24        \\
  \hline
  Fourier, x, 5 cyc./pup.  & 2015-12-16 & -250 \ldots +600 & 35 & CntH        & 1573       & 23        \\
  Fourier, x, 5 cyc./pup.  & 2015-12-16 & -250 \ldots +600 & 35 & BH          & 1625       & 290       \\
  Fourier, x, 5 cyc./pup.  & 2015-12-19 & -250 \ldots +600 & 35 & H$_{2}$     & 2124       & 31        \\
  \hline
  \end{tabular}
\end{table*}

The data were acquired in November and December 2015 during daytime technical time. All measurements were made internally using the light sources available in the calibration unit of the instrument \citep{wildi2009}. Images are taken with the infrared dual-band imager and spectrograph \citep[IRDIS;][]{dohlen2008}, one of the two scientific subsystems of SPHERE dedicated to the detection and characterization of giant planets in the near-infrared \citep{2008A&A...489.1345V,vigan2010}. An exposure time of 10 seconds is used, providing a signal-to-noise ratio high enough for the analysis. The acquired data are summarized in Table~\ref{tab:data_acquisition_log}.

At the beginning of each data acquisition session, the calibration of the sensitivity factor (see Sect.~\ref{sec:calib_aberrations}) was performed to ensure an accurate knowledge of the amount of aberrations introduced in the ZELDA tests. For each test, the PSF was manually centered on the ZELDA phase mask by looking at the intensity image on the IRDIS detector. Since ZELDA transforms phase errors into intensity variations, performing an accurate centering visually is relatively easy, especially since the overall amount of aberrations in SPHERE is low \citep{vigan2016}. The PSF is centered by changing the reference slopes of the near-IR differential tip-tilt sensor (DTTS) of the instrument, the purpose of which is to maintain the PSF centered and stable on the coronagraph. As a result, the tip-tilt cannot be controlled directly with the reference slopes of the SH WFS because the DTTS will ensure that the PSF is always centered on the same location. For the tip-tilt tests with ZELDA, the DTTS loop was opened to disable this control of the centering and allow the PSF to move freely.

\begin{figure*}
  \centering
  \includegraphics[width=1.0\textwidth]{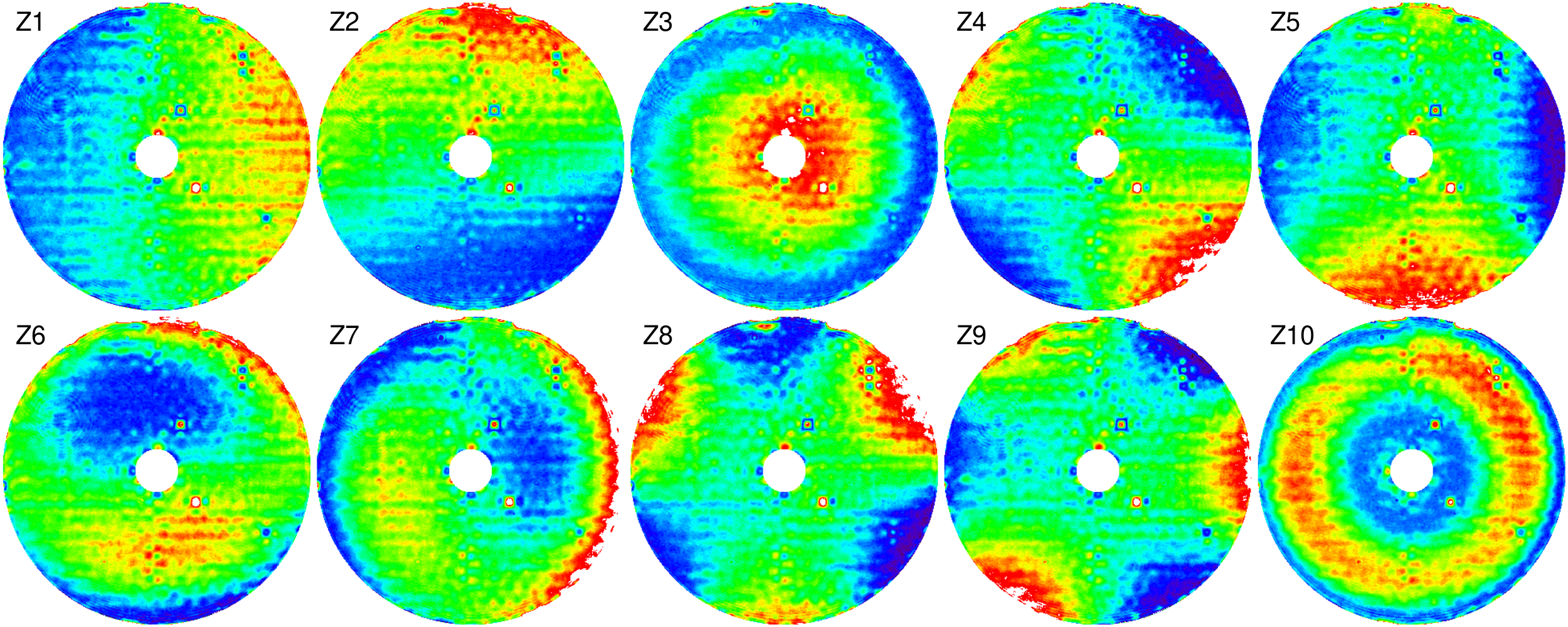}
  \caption{All Zernike modes introduced with 400~nm~PtV and measured with ZELDA. In these maps, the regular pattern of actuators is clearly visible. The dead or stuck actuators of the SPHERE DM appear as white or black circular spots because they are far beyond the linearity range of the sensor. The actuators visible at the edge of the central obscuration (14\% of the pupil in diameter, numerically masked in these maps) are not dead or stuck, but they are not controlled properly because they significantly overlap with the central obscuration.}
  \label{fig:zernike_modes}
\end{figure*}

For each test, three types of data were acquired: ZELDA data with the PSF centered on the phase mask, a clear pupil reference with the PSF outside of the phase mask, and instrumental backgrounds with the same integration time as the ZELDA and clear pupil data. The second type of data gives us the contribution of the amplitude aberrations. All data were acquired in the Fe~{\sc ii} near-infrared narrow-band filter centered on $\lambda=1642$~nm ($\Delta\lambda=24$~nm), for which the geometry of the ZELDA phase mask was originally designed.

The ramp in amplitude was chosen to sample the range within which $\varphi$ is monotonic, which is from -250 to +250 nm peak-to-valley (PtV), and it was extended to +600~nm~PtV to study the nonlinearity observed beyond this range. The range was sampled by 35 measurement points, with a coarse sampling at the extremes and a finer sampling close to zero. All modes were tested with the same ramp. Additional tests were performed with other filters than the Fe~{\sc ii}, see Sect.~\ref{sec:sensitivity_spectral_bandpass}. 

The Zernike modes were calculated on an annular geometry \citep{mahajan1981} to take into account the fact that the central actuators of the SPHERE DM are not seen by the WFS and are not controlled in closed-loop mode by the system. The Fourier modes were tested up to ten cycles/pupil in both horizontal and vertical directions (x and y). At nine cycles/pupil some instabilities of the system started to be observed (high-order AO loop opening, probably due to aliasing effects) for the largest amplitudes (more than 200~nm), therefore the ramp was limited to a smaller extent for these modes.

\subsubsection{Processing}
\label{sec:processing}

After acquisition, the data of each sequence were processed uniformly. The images were first background subtracted, and the bad pixels were corrected for using a sigma-clipping procedure. Then we normalized the ZELDA pupil image by the clear pupil image itself for each pixel, and finally, for each aberration and amplitude, we calculated the phase from the normalized ZELDA images following Eq.~\ref{eq:phase}.

Figure~\ref{fig:zernike_modes} shows all the annular Zernike modes measured with ZELDA in the system when 400~nm~PtV of aberrations are introduced. The central actuators, not seen by the WFS and controlled by the system, have been numerically masked out. Overall, the shape of the Zernike modes is well reproduced and clearly recognizable. The regular pattern of the DM actuators is clearly visible, as is the effect of individual dead or stuck and known actuators.

\subsection{Measurements and comparison with theory}
\label{sec:measurements_comparison_theory}

We analyzed the performance of ZELDA over several spatial frequency regimes by using different Zernike and Fourier modes. For any given mode and amplitude of aberration, we compared the wavefront error measured by ZELDA with the theory based on our numerical model. The consistency between the two was used to validate our concept.

In our tests on SPHERE, the introduced aberration was measured by ZELDA for different error sources: the quasi-static aberrations in our system, dead actuators in the DM, the photon noise and detector readout noise. Since these sources of errors alter the estimate of the introduced aberration, we adopted a differential data analysis strategy to minimize their effect on the ZELDA measurement. For each mode, we first measured the quasi-static errors in the absence of added aberrations on the DM and subtracted this reference from all the other measurements where an aberration is introduced, leaving mostly the introduced aberration, the photon noise and the detector noise. These last two contributors are small thanks to the high signal-to-noise (S/N) ratio of the original images. The dead actuators on the DM generate large amplitude errors that go beyond the dynamic range of ZELDA, making the aberrations at the corresponding pupil locations challenging to estimate. We ruled out the dead actuator points in our measurement map with a numerical mask. From the resulting measurement points, we fit the measured aberration with a function representing the introduced aberration to mitigate the remaining error sources (residual errors, the photon noise and detector noise) and derived the aberration amplitude that was experimentally measured by our Zernike sensor. 

For comparison to the experimental result, we numerically modeled the theoretical ZELDA measurement for the introduced amplitude on DM in the absence of noise. As for experimental data, we excluded the points that are measured at the location of dead actuators. The theoretical measurement map has the same sampling as the experimental map to determine the aberration amplitude given by ZELDA. We finally compared the theoretical and experimental measurements performed by ZELDA for a given mode and amplitude.

\subsubsection{Zernike modes}
\label{subsubsec:Zernike_modes}

\begin{figure}
    \centering
    \includegraphics[width=0.45\textwidth]{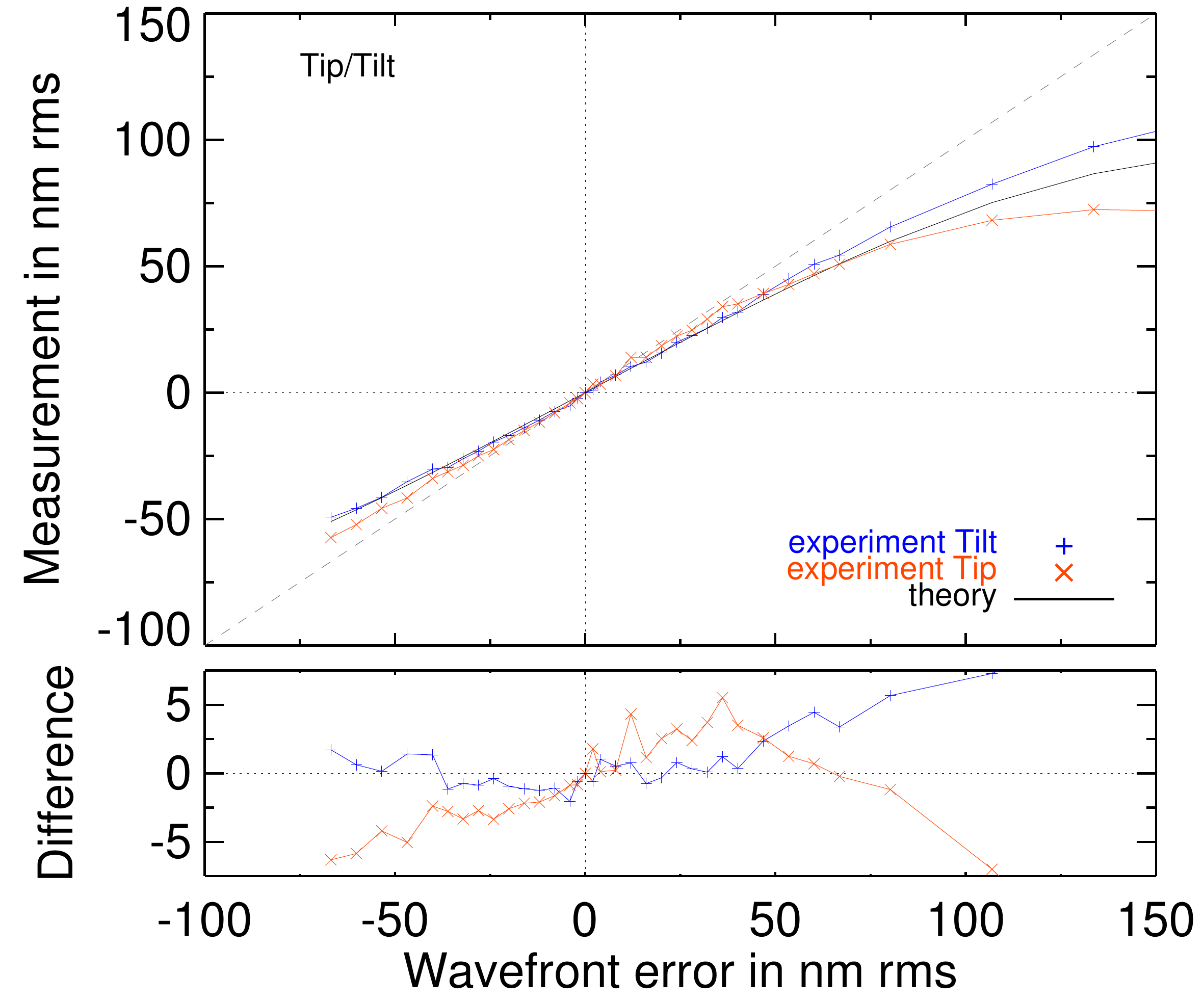}
    \includegraphics[width=0.45\textwidth]{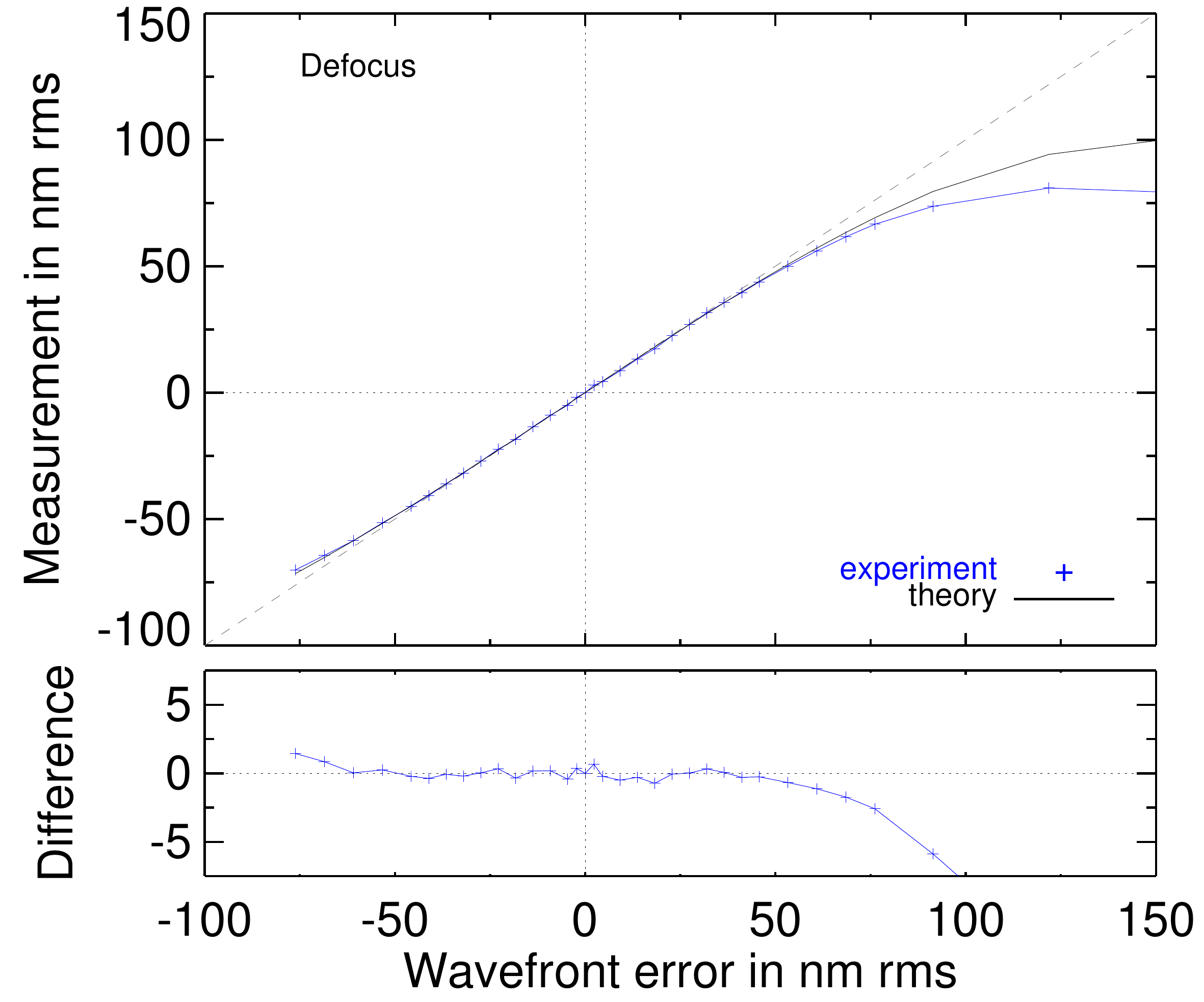}
    \caption{\textbf{Top plot}: Response curves of the ZELDA sensor for tip-tilt (top) and defocus errors (bottom) in simulations (black line) and during the experiment on VLT/SPHERE (colored crosses). \textbf{Bottom plot}: Difference between the experiment and the theory.}
    \label{fig:zernike_mode_low}
\end{figure}

\begin{figure}[!ht]
    \centering
    \includegraphics[width=0.45\textwidth]{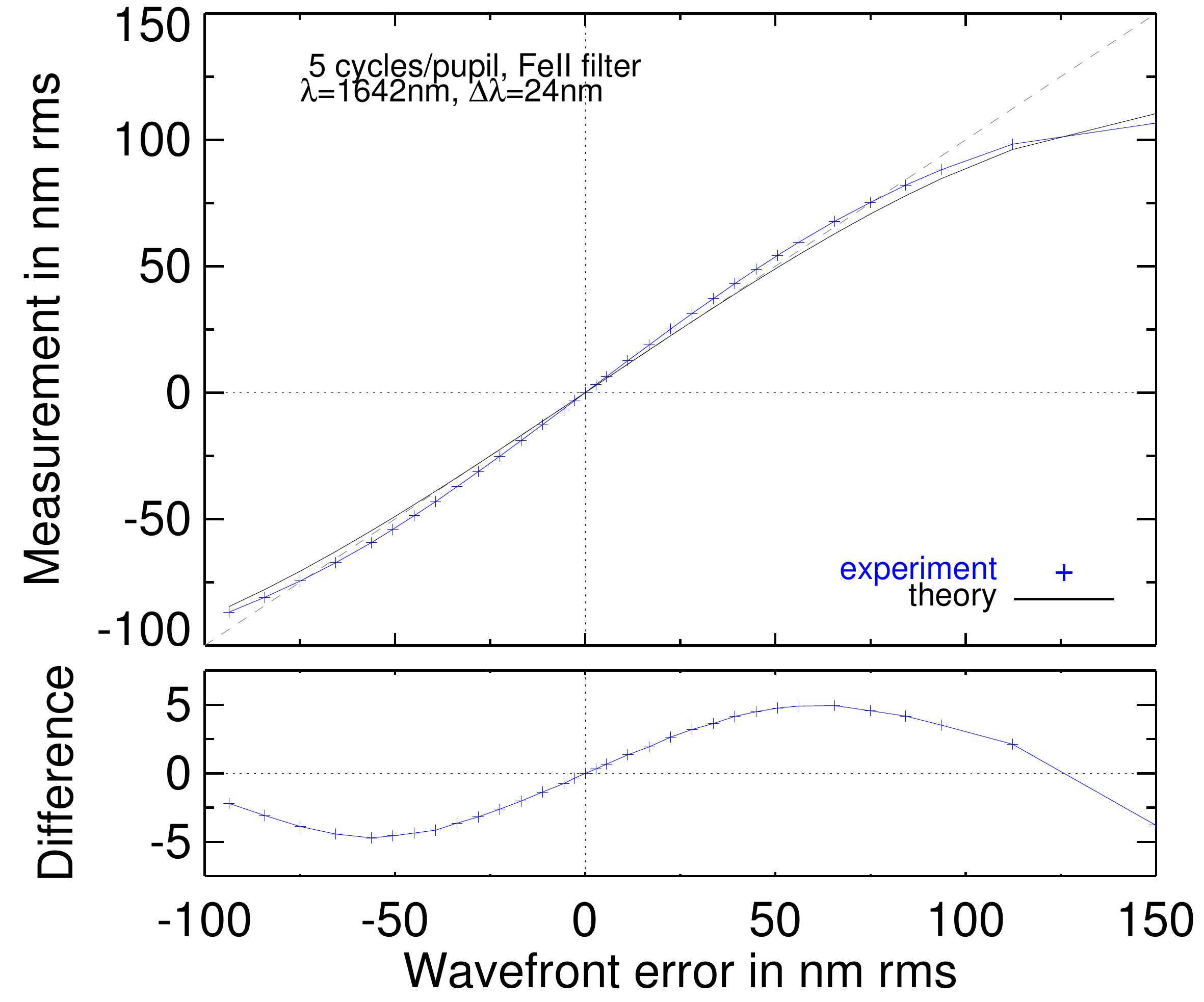}
    \includegraphics[width=0.45\textwidth]{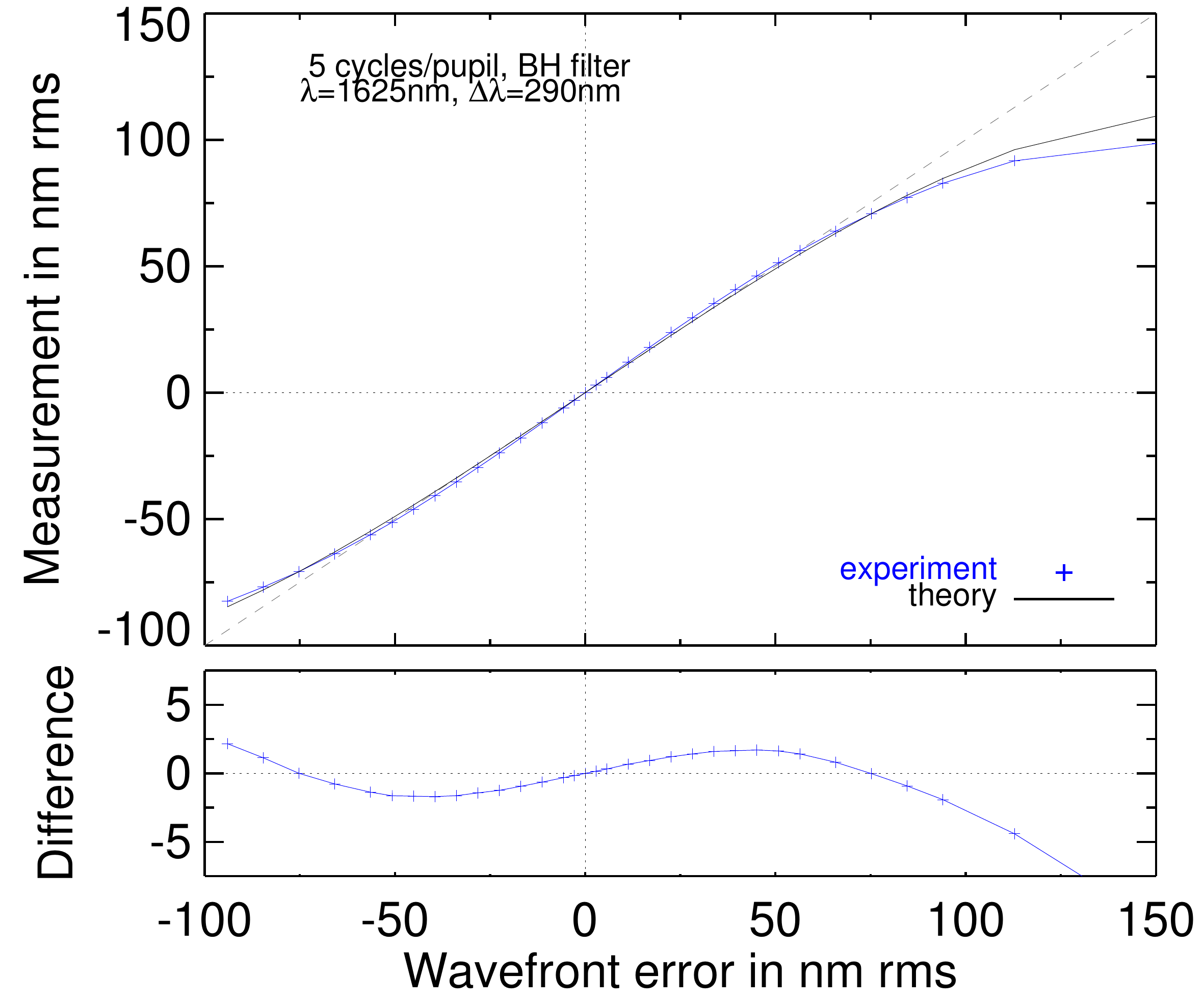}
    \includegraphics[width=0.45\textwidth]{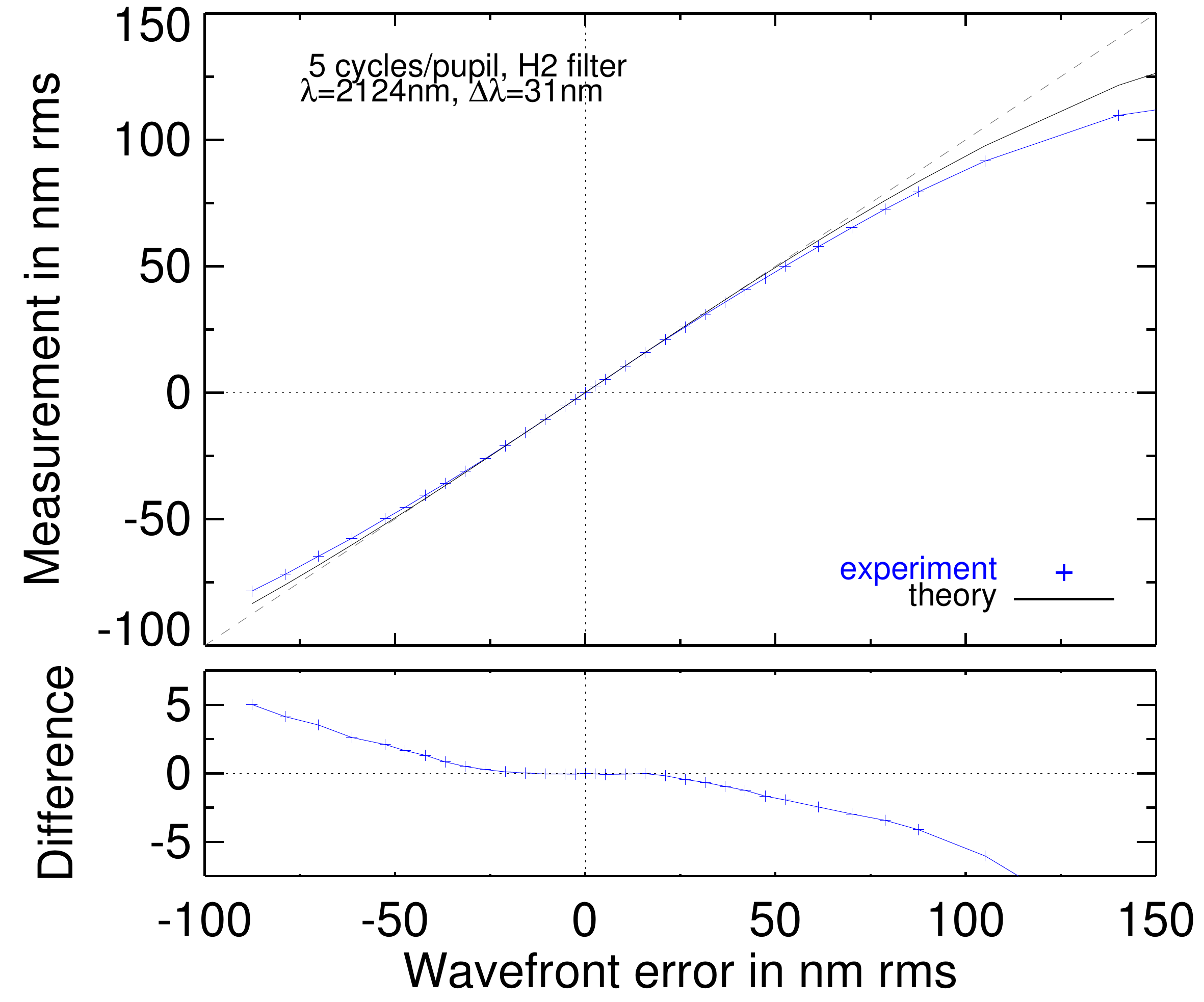}
    \caption{\textbf{Top plot}: Response curves of the ZELDA sensor for the Fourier mode with 5 cycles per pupil along the x-axis direction in simulations (black line) and during the experiment on VLT/SPHERE (blue crosses), using Fe~{\sc ii} (top), BH (middle), and H$_2$ (bottom) filters. The characteristics of the filter are summarized in Table  \ref{tab:data_acquisition_log}. \textbf{Bottom plot}: Difference between the experiment and the theory.}
    \label{fig:chromatism}
\end{figure}

Figure \ref{fig:zernike_mode_low} displays the theoretical and experimental measurements of the Zernike sensor for tip-tilt and defocus in the top and bottom panels, showing good agreement between theory and experience. The same general behavior has been observed with other Zernike modes and Fourier modes, as reported in Appendices \ref{sec:Zernike_modes} and \ref{sec:Fourier_modes}. The discrepancies between theory and experiment are thought to be related to the accurate calibration of the sensitivity factor. We note that the value of the sensitivity factor varies by up to 10\% from one day to the next. No systematic measurements were made to quantify the variations at higher temporal frequency, but it is reasonable to assume that some variations can be expected over the course of 5-6 hours (typical length of our data acquisition sessions), leading to the small differences observed with respect to the theory.

The Zernike sensor shows a linear response around the zero point, enabling a simple reconstruction of the small coronagraphic aberrations and fast convergence with close-loop compensation toward the zero point in one or two iterations. In this regime, the slope of the response curve is equal to one for all the Zernike modes except for the tip-tilt errors. As underlined in \citetalias{N'Diaye2013a}, this behavior is believed to be due to the modification of the light distribution going through the decentered mask with respect to the star image. With a careful calibration, this effect will have no impact on the ability of the Zernike sensor to measure tip-tilt errors.

For the aberrations outside the linear range, the Zernike sensor still proves efficient by considering a close-loop operation between with the measurement and correction. The measurements converge towards the linear regime of the wavefront sensor after a few iterations, as shown by \citet{Vigan2011}. Since ZELDA is assumed to operate in the presence of small aberrations, our data analysis is based on a quadratic relation between the measured intensity and the wavefront errors (see Sect. \ref{sec:ZELDA_sensor}). Investigating a more accurate reconstruction expression could increase the measurement accuracy for large phase errors but this is beyond the scope of the paper.

\subsubsection{Sensitivity to spectral bandpass}
\label{sec:sensitivity_spectral_bandpass}

The above results have been given for an internal light source passing through a Fe~{\sc ii} narrow-band filter of central wavelength $\lambda_c=$1642\,nm and bandwidth $\Delta\lambda=$24\,nm, for which the ZELDA sensor has been originally designed. In \citetalias{N'Diaye2013a} we numerically studied the sensitivity of the Zernike sensor to chromatic effects, showing the reliability of our concept for a light source with well-known central wavelength. We now probe the performance of our sensor experimentally by measuring aberrations in light source filtered with different chromatic parameters on SPHERE.

We analyzed the ZELDA response curve for a Fourier aberration mode with five cycles per pupil in three different filters: Fe~{\sc ii} ($\lambda_c$=1642\,nm, $\Delta\lambda$=24\,nm), BH ($\lambda_c$=1625\,nm, $\Delta\lambda$=290\,nm), and H$_{2}$ ($\lambda_c$=2124\,nm, $\Delta\lambda$=31\,nm). Based on the nature of the considered filters, we estimated the introduced aberration with a data analysis based on the terms $\theta$ and $b$ that were re-evaluated at $\lambda=\lambda_c$, as shown in Fig. \ref{fig:zelda_b-profile}. The ZELDA measurements for these three filters are given in Fig.~\ref{fig:chromatism}.

For each filter, the experimental measurements agree well with the numerical values. The match in BH and H$_2$ filters seems slightly better than in Fe~{\sc ii} filter, and we see two reasons that might explain this result: first of all, a better calibration of the sensitivity factor when the test with BH and H$_2$ filters were performed, and second, the smaller amount of aberrations with respect to the wavelength at redder filters, enabling a more accurate measurement of the small errors by ZELDA, provided that the sensor has enough dynamic range to operate.

More interestingly, the second and third panels underline the ability of ZELDA to perform accurate aberration measurements using filters with a wide spectral band ($\Delta\lambda$=290\,nm) or centered on a wavelength largely shifted from the wavelength of design ($\lambda_c=$2124\,nm instead of $\lambda_0=$1642\,nm, i.e., a shift of 482\,nm), as long as the data analysis accounts for the filter characteristics in its reduction. This interesting property of our sensor could prove beneficial for the estimation of chromatic aberrations by operating multichromatic aberration measurements with ZELDA at several narrow-band filters, ideally in parallel. Such measurements could possibly lead to contrast gain in post-processing methods based on spectral deconvolution \citep{2002ApJ...578..543S,2008A&A...489.1345V} and hence, improvement in the extraction of the spectral information for the faintest planetary companions. 

\section{Results with ZELDA-based wavefront correction}

\begin{figure}
  \centering
  \includegraphics[width=0.5\textwidth]{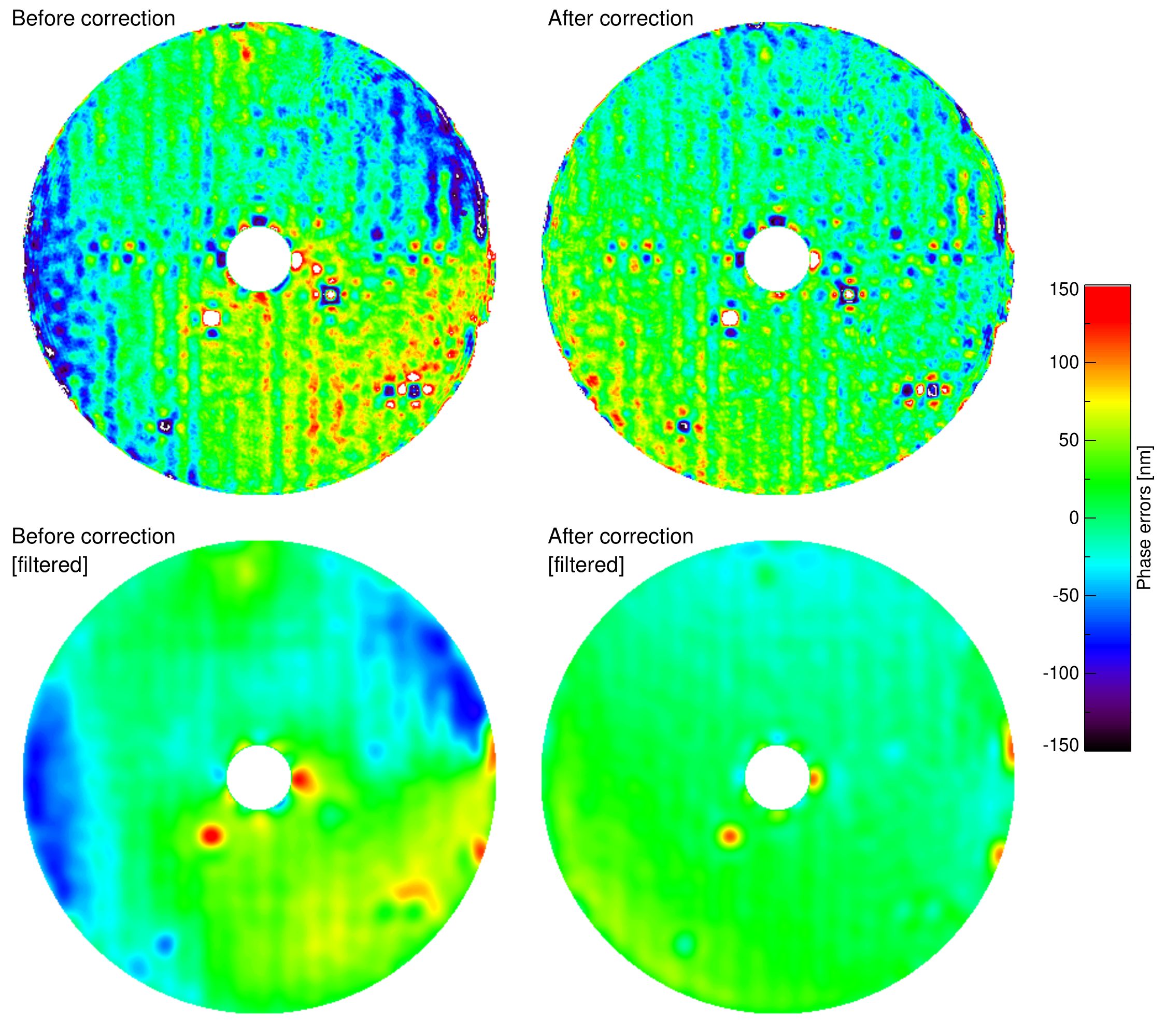}
  \includegraphics[width=0.5\textwidth]{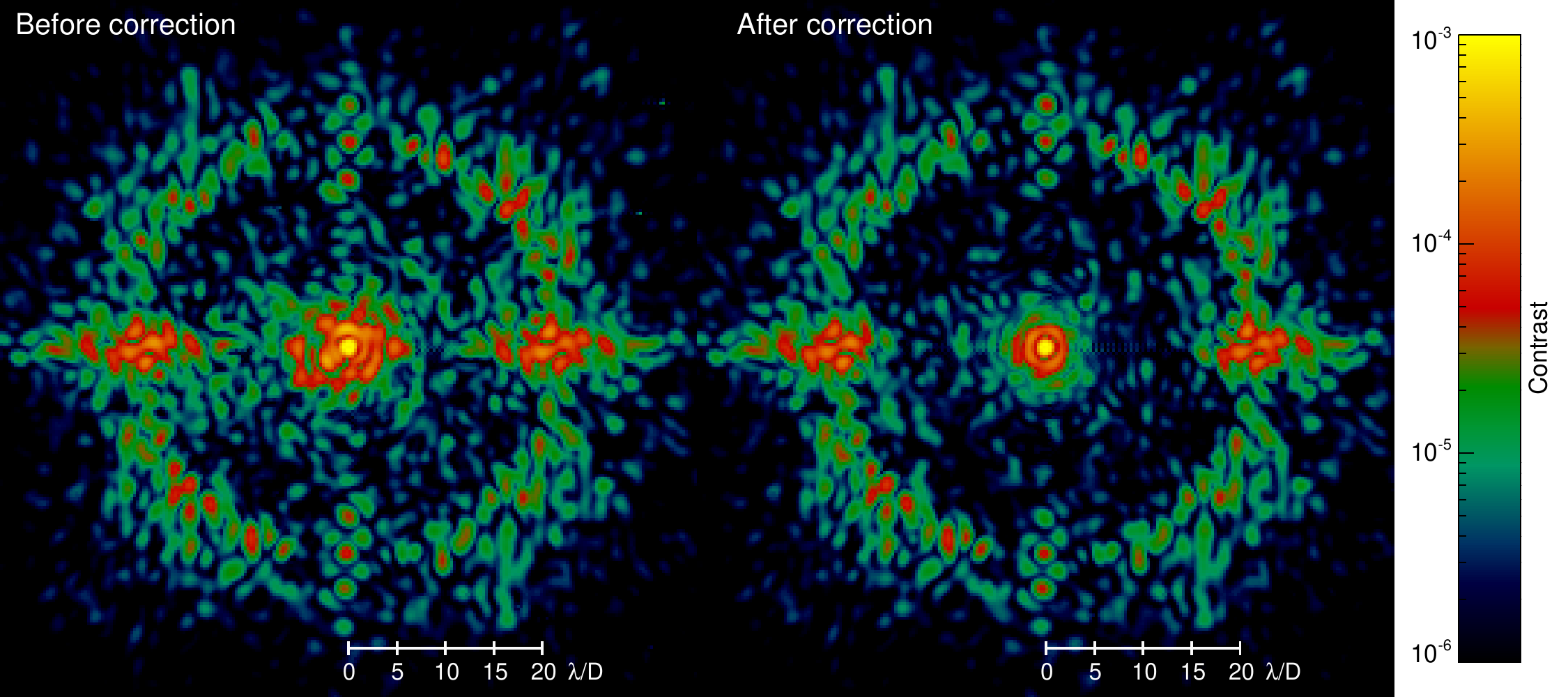}
  \caption{Illustration of the correction of the SPHERE NCPA using ZELDA measurements with IRDIS. The top row shows the OPD maps measured with ZELDA before (left) and after (right) the correction, presented at the same color scale. The middle row shows the same maps filtered in Fourier space using a Hann window of size 25~\lsd (see text). Low spatial frequency aberrations are clearly visible on the left, while they have disappeared after the correction, except for a small amount of residual tip-tilt (see text). The bottom row shows the equivalent focal plane coronagraphic images before (left) and after (right) the compensation of the NCPA, measured at 1593~nm (IRDIS H2 filter) and presented at the same color scale. The gain is obvious close to the center, but also noticeable farther out where the intensity of the speckles in the corrected area has decreased significantly.}
  \label{fig:full_projection_opd_coro}
\end{figure}

In this section, we now use ZELDA for the purpose it was originally designed for: measuring and compensating for the NCPA of SPHERE at the level of the coronagraphic mask to improve the quality of the focal-plane images. First, the amount of aberrations introduced when changing the reference slopes was calibrated. Then a ZELDA measurement was acquired and analyzed to produce an OPD map, which is presented in the top left of Fig.~\ref{fig:full_projection_opd_coro}. In addition to the static pattern of DM actuators, the OPD map clearly shows low spatial frequency aberrations at the level of a few dozen nanometers (RMS), corresponding to uncorrected NCPA.

This map cannot be projected directly onto the reference slopes of the WFS. The sampling of the pupil on the IRDIS detector is such that ZELDA is sensitive to spatial frequencies higher than 190 cycles/pupil, while the SPHERE high-order DM has only 40$\times$40 actuators, which in theory allows corrections of up to 20~\lsd. To avoid any spatial aliasing, the ZELDA OPD map was first filtered in Fourier space using a Hann window of size 25~\lsd\xspace (Fig.~\ref{fig:full_projection_opd_coro}, middle row). We used a 25~\lsd window because it provides a better correction than a 20~\lsd. The reason is that because the Hann window falls exactly at zero at the window edge, spatial frequencies that could be corrected with the DM are too strongly attenuated or even completely filtered out in the process. In our limited time with the instrument, we did not have the opportunity to investigate further than testing different window sizes. The use of other filtering windows may provide better results.

\begin{figure}
  \centering
  \includegraphics[width=0.5\textwidth]{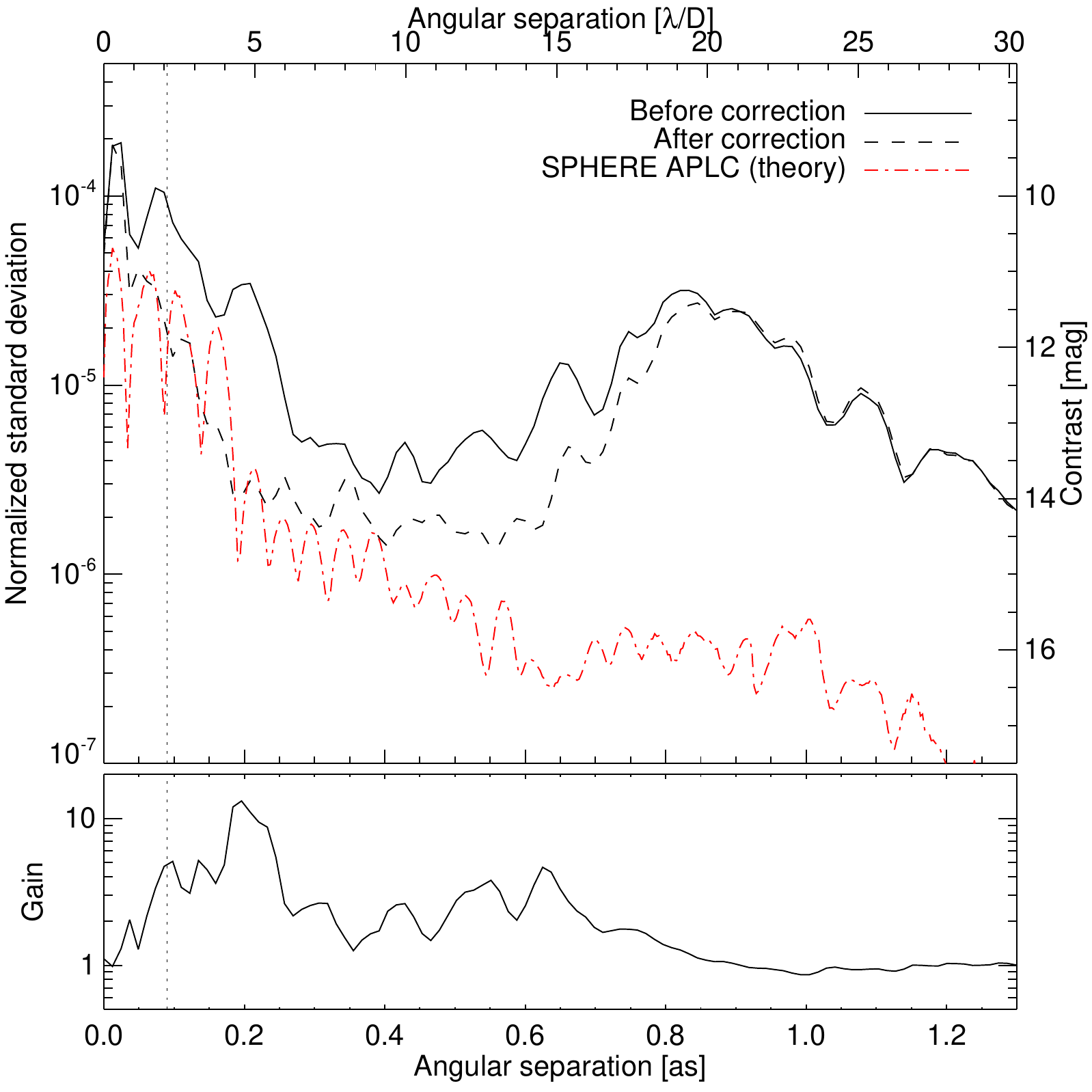}
  \caption{Normalized azimuthal standard deviation profiles before (plain line) and after (dashed line) correction of the NCPA using ZELDA, as a function of angular separation. The contrast gain is plotted in the bottom panel. The dotted line corresponds to the edge of the coronagraphic mask (90 mas). The measurements correspond to the coronagraphic images presented in Fig.~\ref{fig:full_projection_opd_coro}. They are compared to the theoretical performance of the SPHERE APLC (red, dash-dotted line).}.
  \label{fig:full_projection_contrast}
\end{figure}

The filtered OPD map was then projected onto the reference slopes of the WFS, and a new ZELDA image was acquired and analysed, providing the OPD map showed in Fig.~\ref{fig:full_projection_opd_coro}, top right panel. In this new OPD map, the low spatial frequencies that were clearly visible before the correction have now completely disappeared, and the map appears much flatter. The OPD maps show that the NCPA of the system apparently has indeed been corrected. A slight residual tip-tilt in the OPD map after correction remains, particularly visible in the Hann-filtered version of the OPD map. This is not the result of an imperfect measure, but simply arises because the tip-tilt is controlled in close-loop mode by the DTTS (see Sect.~\ref{sec:data_acquisition_sphere}), for which the reference slopes are changed manually at the beginning of the test to center the PSF on the ZELDA mask. The optimal approach would be to correct the high orders by modifying the reference slopes of the SH WFS and to correct the tip-tilt by modifying the reference slopes of the DTTS. We were unable to investigate this approach for the tests presented here, but it will represent a necessary improvement in future tests.

To verify the quality of the NCPA compensation, we acquired coronagraphic images with IRDIS at 1593~nm (H2 filter, see, e.g., \citet{vigan2010}), using the Apodized Pupil Lyot Coronagraph \citep[APLC;][]{Soummer2005} optimized for the $H$ band. Two data sets were acquired: one using the default reference slopes of the WFS (before correction), and one using the reference slopes updated using the ZELDA measurement (after correction). For each data set, we acquired a 2 min coronagraphic image and 2 min reference PSF image where the PSF was moved out of the coronagraphic mask. A neutral density filter was used to acquire the reference PSF without saturating the peak. Corresponding instrumental backgrounds were also acquired. The resulting coronagraphic images are showed in the bottom row of Fig.~\ref{fig:full_projection_opd_coro}. The visual difference between the two images is striking inside the AO-corrected area. The image before correction is dominated by speckles close to the axis, up to 7-8~\lsd. It also shows a strong horizontal and vertical pattern of speckles that extend from the edge of the corrected region down to $\sim$10~\lsd. After the NCPA correction, the whole AO-corrected region appears much cleaner: the speckles close to the axis have almost disappeared to reveal a very regular annulus at the edge of the coronagraphic mask, similar to what is expected from the theoretical design of the SPHERE APLC \citep{guerri2011}. In the 4-8~\lsd range, the static speckles are also strongly attenuated. In the remaining AO-corrected region, the speckles are also attenuated, which is particularly visible along the horizontal and vertical directions, where a strong static pattern of speckles was previously visible.

To quantitatively assess the performance gain after NCPA compensation, we plot in Fig.~\ref{fig:full_projection_contrast} the azimuthal standard deviation of the coronagraphic images as a function of separation, normalized to the peak flux of the reference off-axis PSF. The bottom panel of the figure shows the gain in contrast between the two curves. Within 2-16~\lsd, there is a gain in contrast of a factor more than 2, with even a peak at more than 10 around 5~\lsd. This agrees very well with estimates from \citetalias{N'Diaye2013a}, where we estimated a possible gain over a factor of 10, and it is a strong confirmation of the potential of ZELDA to compensate for NCPA. We also plot the simulated theoretical performance of the SPHERE APLC in presence of the amplitude aberrations directly measured in the instrument and using an image of the Lyot stop, but without any phase aberrations. Within 5~\lsd, we reach this theoretical performance, which means that the NCPA at low spatial frequencies are almost entirely corrected for. We note that on the internal source the instrument pupil is purely circular, with no central obscuration or spiders, but the Lyot stop still includes elements to mask the diffraction of the central obscuration, the spiders, and the bad actuators of the DM. This will result in a slightly better performance of the coronagraph on sky than on the internal source.

\section{Conclusion}

In \citetalias{N'Diaye2013a}, we proposed a Zernike phase-mask sensor for the measurement of coronagraphic aberrations in exoplanet direct-imaging instruments. In this paper, we have presented the first experimental results and wavefront error correction with our concept in a real instrument, allowing us to improve the image quality in the coronagraphic images for exoplanet observation.

We designed and manufactured a prototype called ZELDA, which was installed inside VLT/SPHERE during its reintegration in Paranal in 2014. We validated the concept experimentally with tests performed on the internal point source where we manually introduced Zernike and Fourier modes on the DM. Following these encouraging results, we measured and compensated for the long-lived NCPA on SPHERE using the measurements from ZELDA. A contrast gain of up to one order of magnitude was reached at an angular separation of 5\,$\lambda/D$ from the axis in the coronagraphic images, providing very encouraging results for the observation of the giant gaseous planets down to the raw contrast limit set by the coronagraph. These performance results are at least on the same order as the contrast gain achieved by \citet{Paul2014} using the coronagraphic phase-diversity method named COFFEE. Additional studies will be performed to fairly compare the NCPA measurements and the contrast performance obtained by ZELDA and COFFEE.

While our first results are very encouraging, we have identified ways to improve the NCPA correction and thus the contrast performance. Our NCPA correction here relied on a ZELDA OPD map that was filtered in Fourier space with a Hann window as a first attempt. The quality of our wavefront error compensation is expected to be even more improved by investigating different sizes and natures of filter windows for the Zernike sensor map. Time constraints made our approach for the ZELDA-based wavefront calibration unsatisfactory for the correction of tip-tilt and higher-order modes because we did not account for all the specificities of SPHERE. Additional tests will enable us to take advantage of all the control loops in the instrument and achieve an optimal correction of the NCPA. We will also investigate combinations of ZELDA with IFS or polarimetric imager to calibrate chromatic or polarimetric aberrations. 

Our experiment was performed with an internal source during our run on SPHERE, potentially representing a first step toward implementing a ZELDA sensing path on the instrument to improve its high-contrast observations on-sky. First phase-error maps have been obtained on sky with our concept to diagnose LWE observed on the coronagraphic images at wind speed below 1\,m/s \citep{Sauvage2016}. Additional thorough studies will lead us to operate on-sky observations with ZELDA-based wavefront calibration and reach of the ultimate contrast limits of the instrument. 

From a practical point of view, the NCPA correction with ZELDA in SPHERE could be implemented following two approaches. The \emph{off-line} approach, in which the NCPA would be measured at the beginning of the night, and the same correction would be applied to all observations during that night. This assumes that most of the NCPA do not vary significantly over the course of a few hours. This is similar to the original calibration scheme foreseen for the instrument, where the NCPA would be calibrated once per day using phase diversity techniques \citep{sauvage2007}. The \emph{online} approach is more complex to implement since it requires replacing the entire near-infrared DTTS with a ZELDA-based sensor. The main gain of the \emph{online} implementation is that it allows sensing not only tip-tilt variations but also higher-order NCPA in real-time during the observations, providing the equivalent of the calibration wavefront sensor in the Gemini/GPI instrument \citep{2010SPIE.7736E.179W}. This solution is currently under study for future evolutions of SPHERE.

The design simplicity and fast algorithm speed of our Zernike sensor for phase reconstruction makes its use very appealing for online measurements of the quasi-static coronagraphic aberrations during observations in various instruments. Given its properties, such a concept has already been adopted as a low-order wavefront sensor for the WFIRST mission and its coronagraphic module \citep{Zhao2014,Spergel2015}. Current high-contrast imaging facilities and future exoplanet direct imagers on ELTs or envisioned post-JWST space observatories \citep{Dalcanton2015} might thus benefit from the ZELDA-based wavefront correction to increase the signal-to-noise ratio of the planetary companions in the coronagraphic images and hence expand the discovery space of observable exoplanets by reaching deeper contrasts at small angular separations.

\begin{acknowledgements}
The authors are very grateful to the referee, Frans Snik, for his insightful suggestions and helpful comments to improve the quality of the original manuscript.

This work is partially supported by the National Aeronautics and Space Administration under Grants NNX12AG05G and NNX14AD33G issued through the Astrophysics Research and Analysis (APRA) program (PI: R. Soummer).  MN would like to thank Rémi Soummer and Laurent Pueyo for their support. MN would also like to acknowledge the ESO Chile Visiting Scientist program. Finally, AV would like to thank the Paranal staff for their patience and support when performing the tests presented in this work.

\smallskip \\

SPHERE is an instrument designed and built by a consortium consisting of IPAG (Grenoble, France), MPIA (Heidelberg, Germany), LAM (Marseille, France), LESIA (Paris, France), Laboratoire Lagrange (Nice, France), INAF - Osservatorio di Padova (Italy), Observatoire de Genève (Switzerland), ETH Zurich (Switzerland), NOVA (Netherlands), ONERA (France) and ASTRON (Netherlands) in collaboration with ESO. SPHERE was funded by ESO, with additional contributions from CNRS (France), MPIA (Germany), INAF (Italy), FINES (Switzerland) and NOVA (Netherlands). SPHERE also received funding from the European Commission Sixth and Seventh Framework Programmes as part of the Optical Infrared Coordination Network for Astronomy (OPTICON) under grant number RII3-Ct-2004-001566 for FP6 (2004-2008), grant number 226604 for FP7 (2009-2012) and grant number 312430 for FP7 (2013-2016).

\end{acknowledgements}

\appendix

\section{Zernike modes}\label{sec:Zernike_modes}

\begin{figure*}
\centering
\resizebox{\hsize}{!}{
\includegraphics{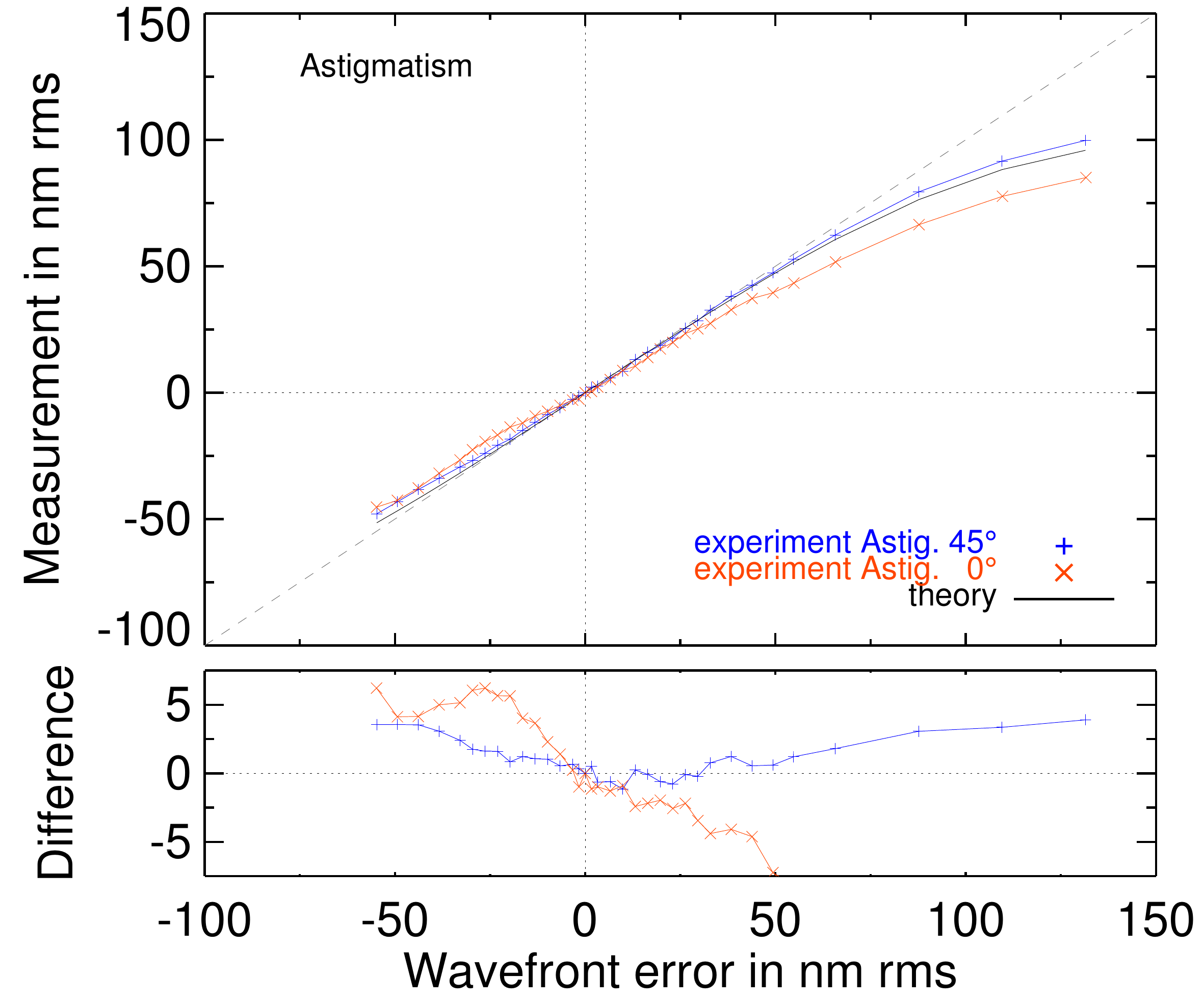}
\includegraphics{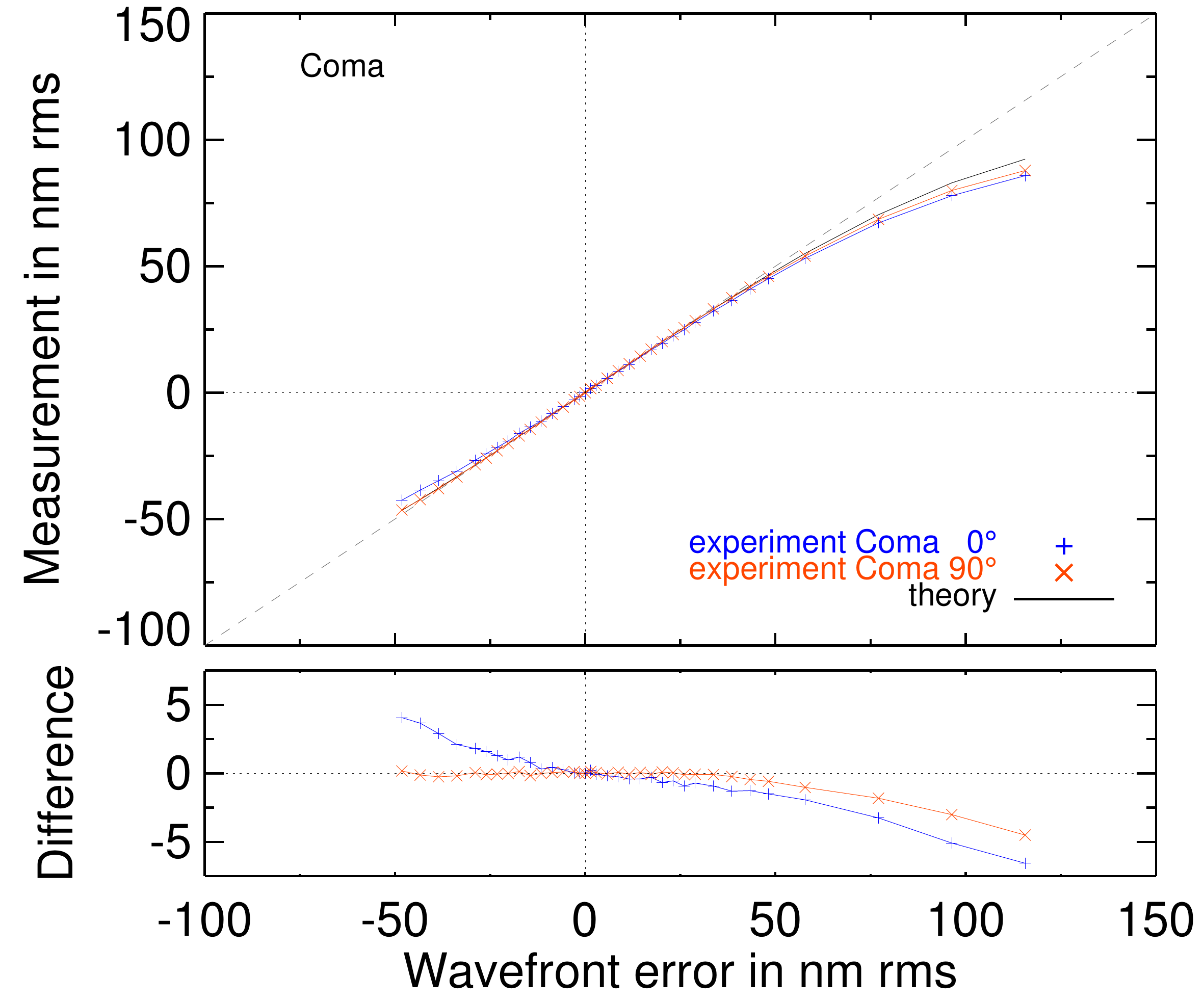}
}
\resizebox{\hsize}{!}{
\includegraphics{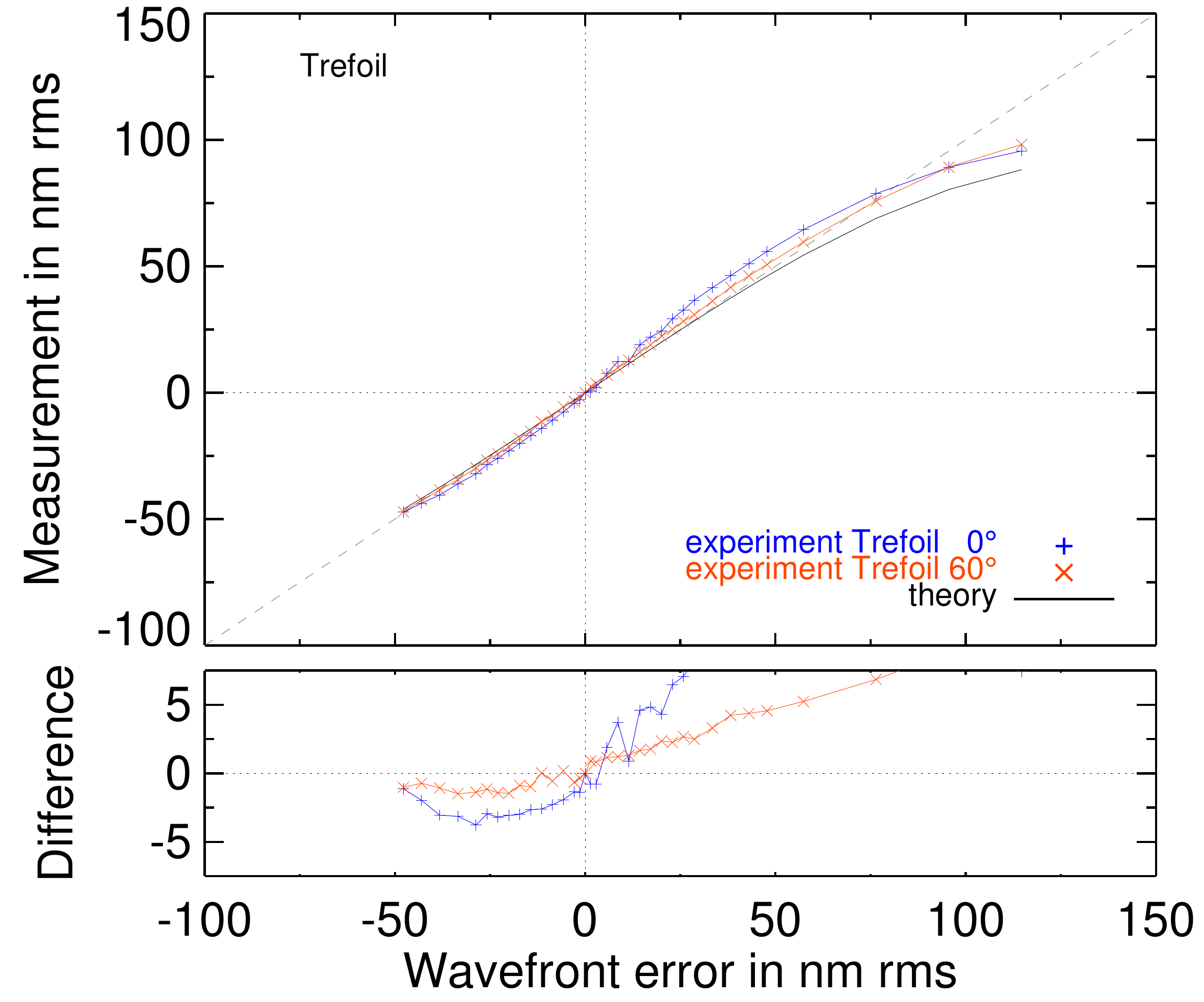}
\includegraphics{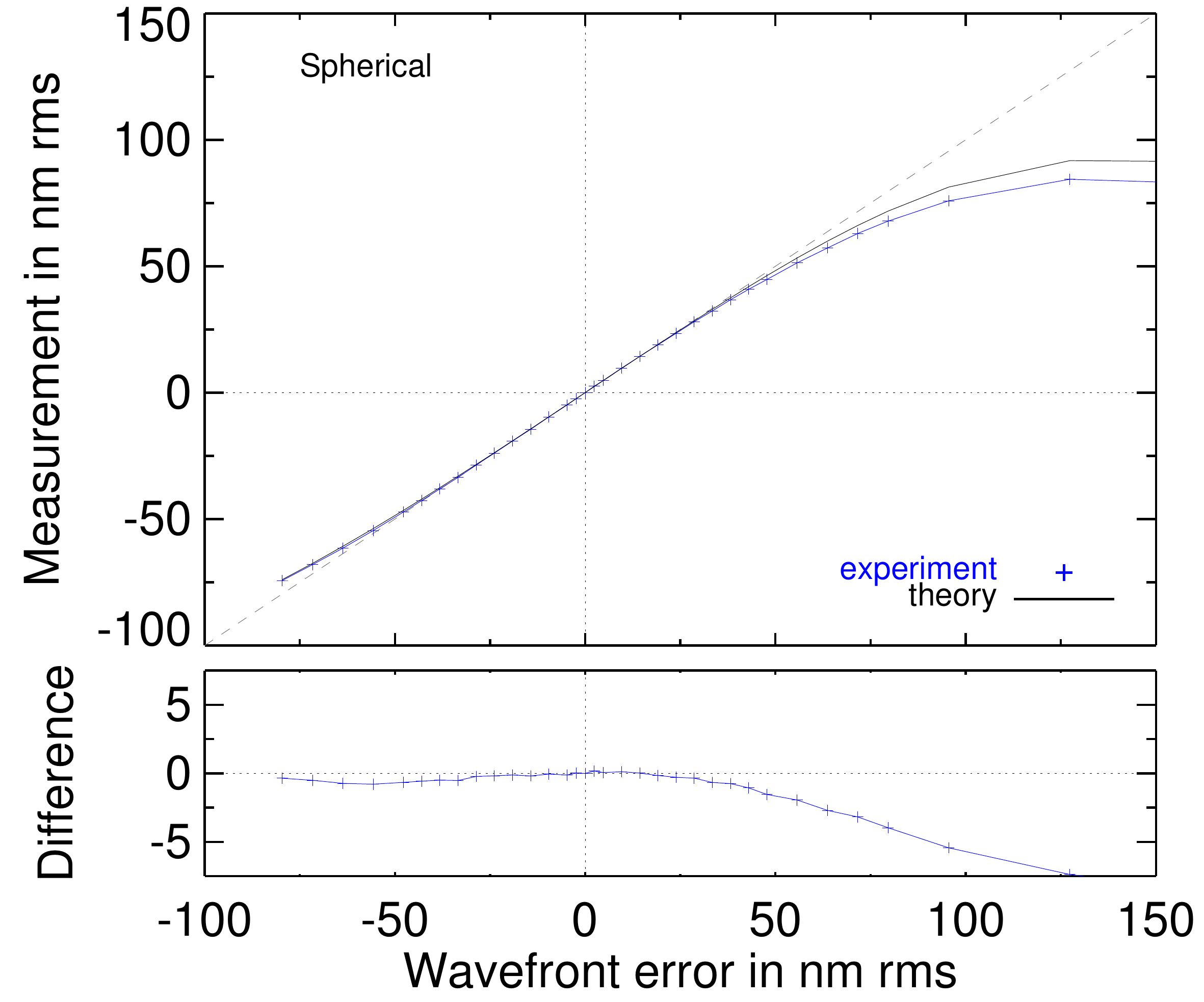}
}
\caption{\textbf{Top plot}: Theoretical and experimental aberration measurements for different Zernike modes as a function of the wavefront error in the entrance pupil plane.\textbf{Bottom plot}: Difference between the experiment and the theory.}
\label{fig:zernike_mode}
\end{figure*}

During our tests on SPHERE, we measured not only the response curves of ZELDA for tip-tilt and defocus modes (see Sect.~\ref{subsubsec:Zernike_modes}), but also for high-order modes: astigmatism, coma, trefoil, and spherical aberration. Figure \ref{fig:zernike_mode} display the results for these modes, showing good consistency between the simulated and experimental measurements. The discrepancies observed between theory and experiment for the tip-tilt error and defocus are possibly due to the imperfect calibration of the sensitivity factor. Our sensor enables the measurement of Zernike modes with nanometric accuracy, which is interesting for realignment purposes and online compensation of aberrations that are due to thermal or optomechanical drifts of the optics for an exoplanet direct-imaging instrument.

\section{Fourier modes}\label{sec:Fourier_modes}

We have also analyzed the response of ZELDA to Fourier modes with different spatial frequencies to assess the ability of our concept to capture speckles at different angular separations from the star position in the coronagraphic images. Section~\ref{subsubsec:Zernike_modes} shows the results obtained by ZELDA for the Fourier mode with a spatial frequency of five cycles per pupil in different filters. In addition to these tests, we also perform systematic ZELDA measurements of the Fourier modes with spatial frequencies ranging from two to ten cycles per pupil in both x- and y-axis directions, but using only the Fe~{\sc ii} filter.

\begin{figure*}
\centering
\resizebox{\hsize}{!}{
\includegraphics{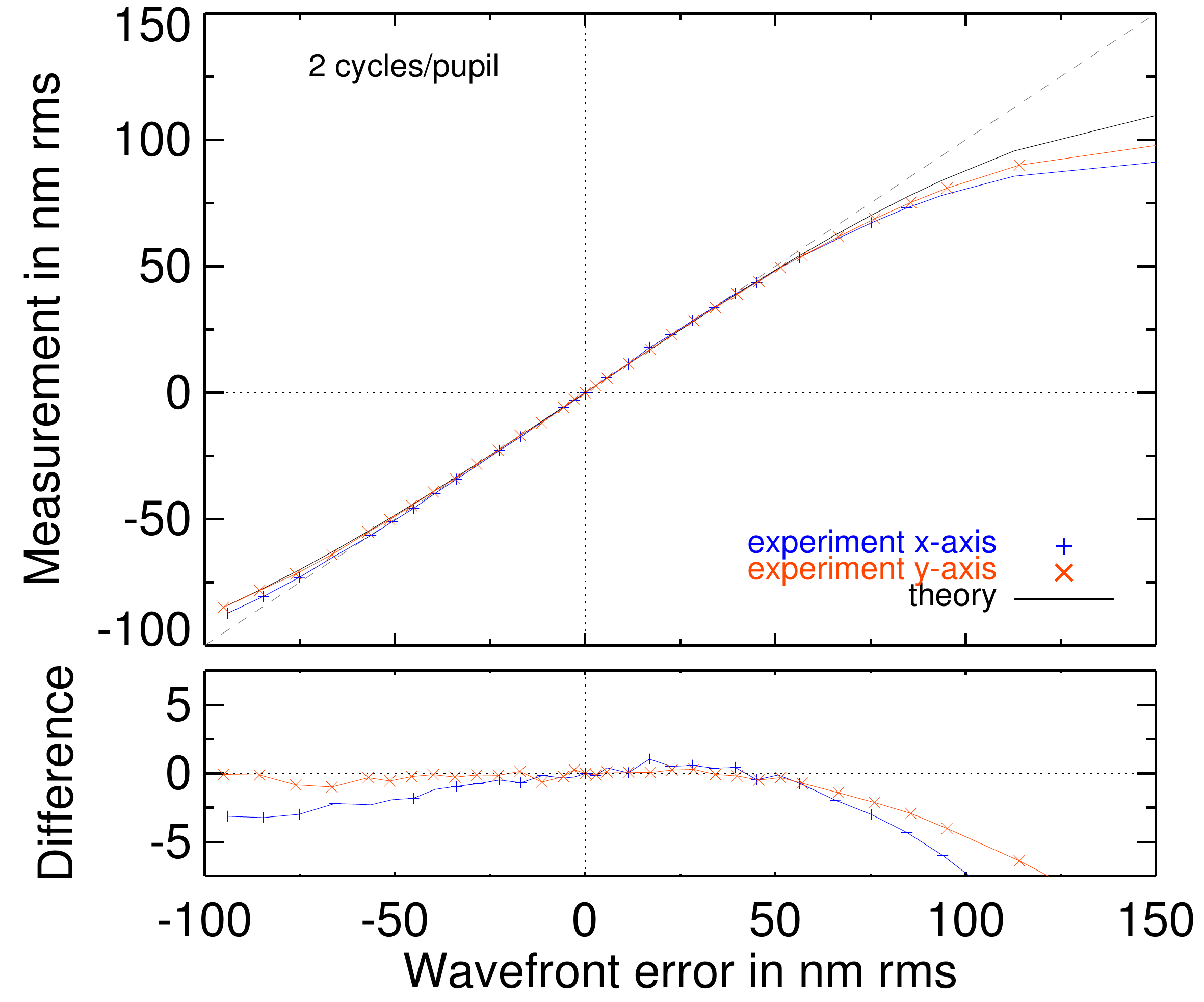}
\includegraphics{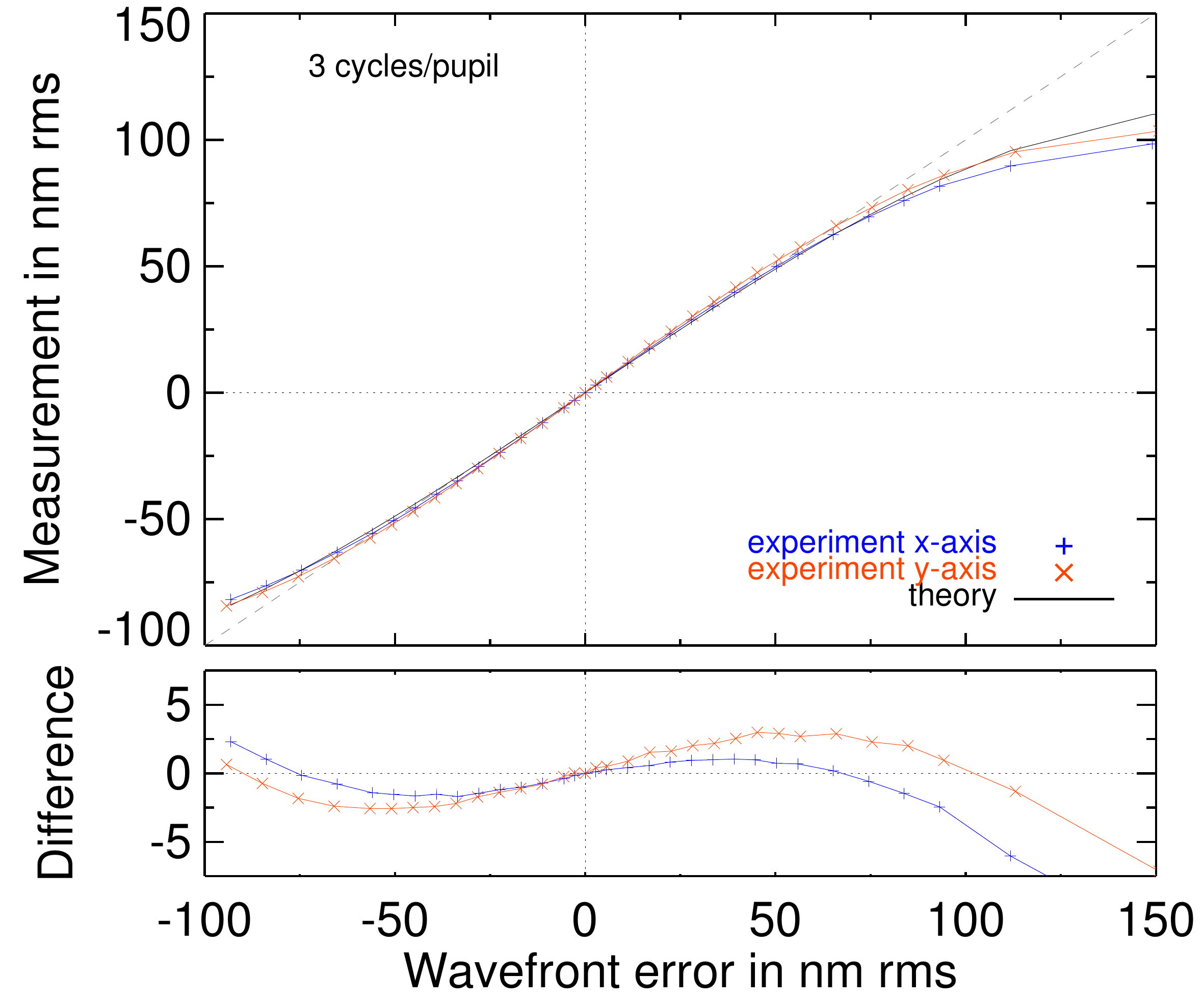}
\includegraphics{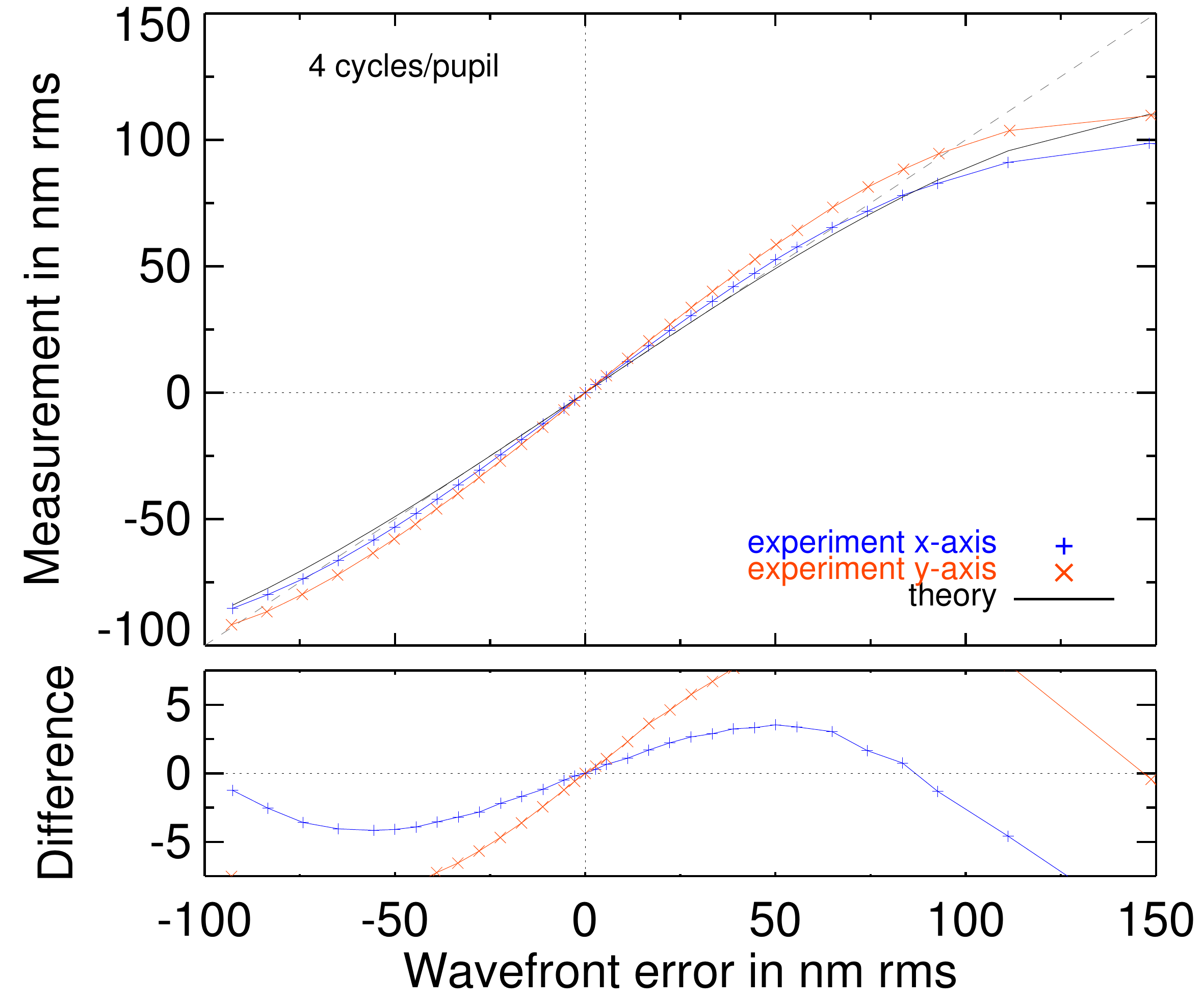}
}
\resizebox{\hsize}{!}{
\includegraphics{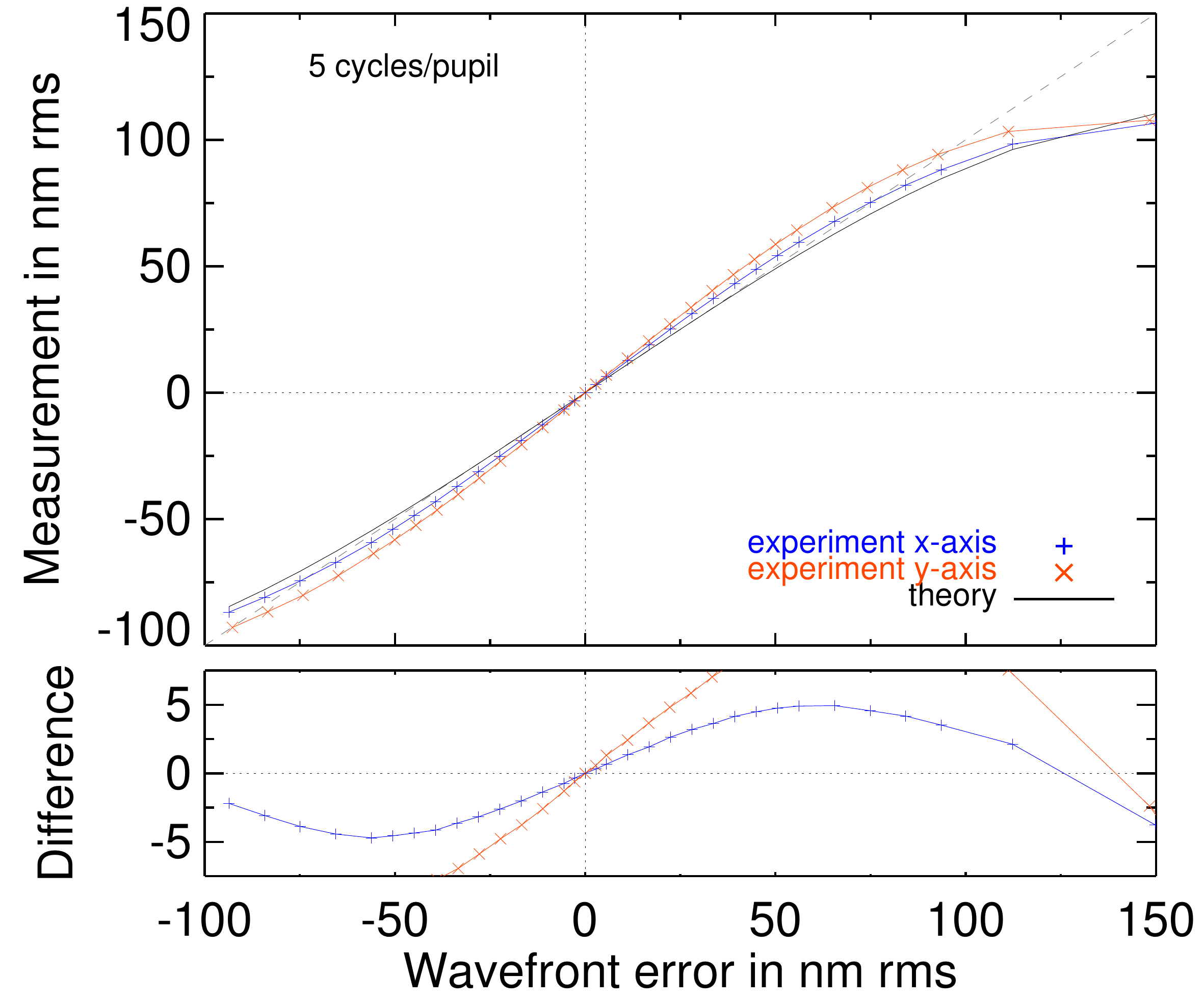}
\includegraphics{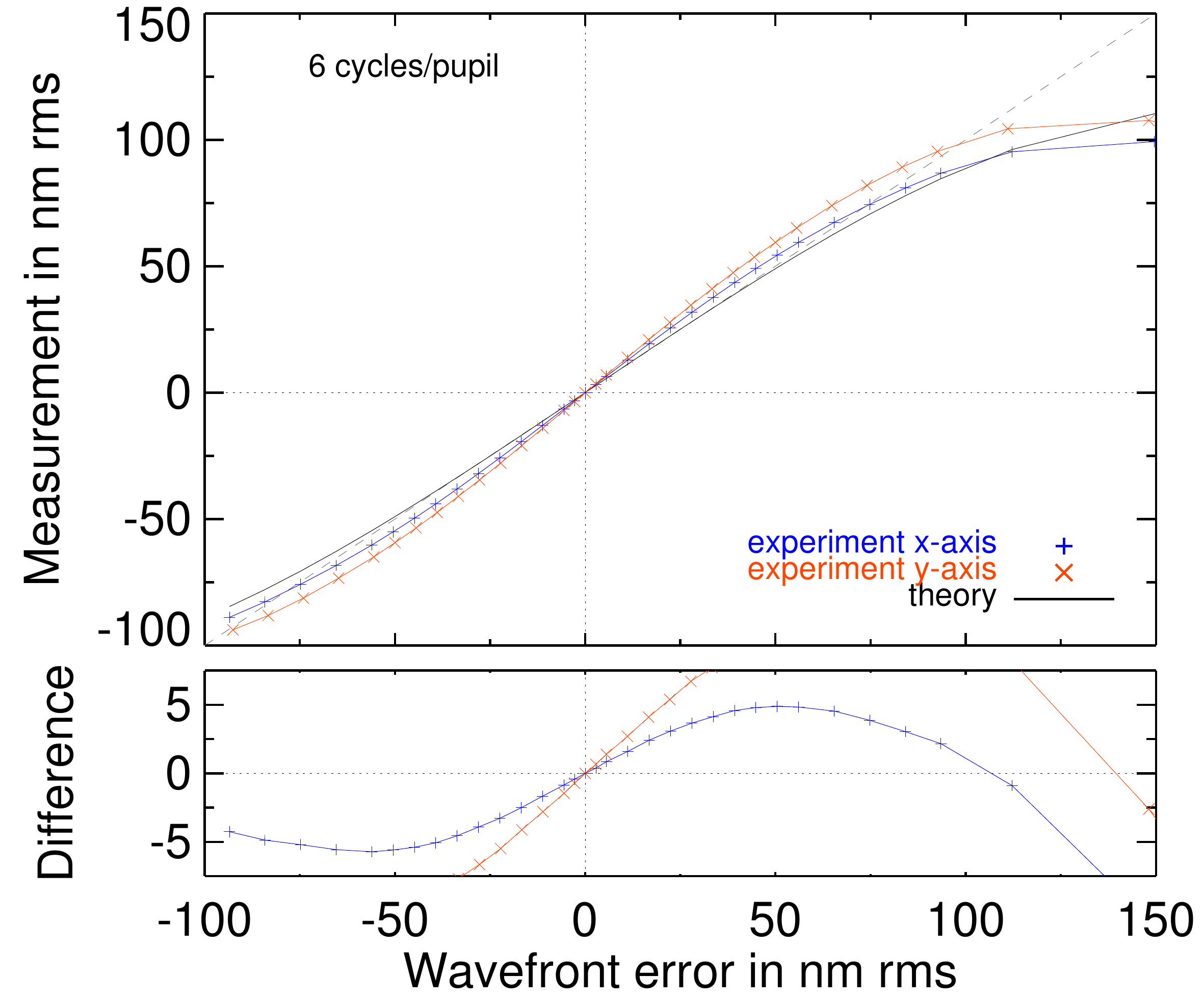}
\includegraphics{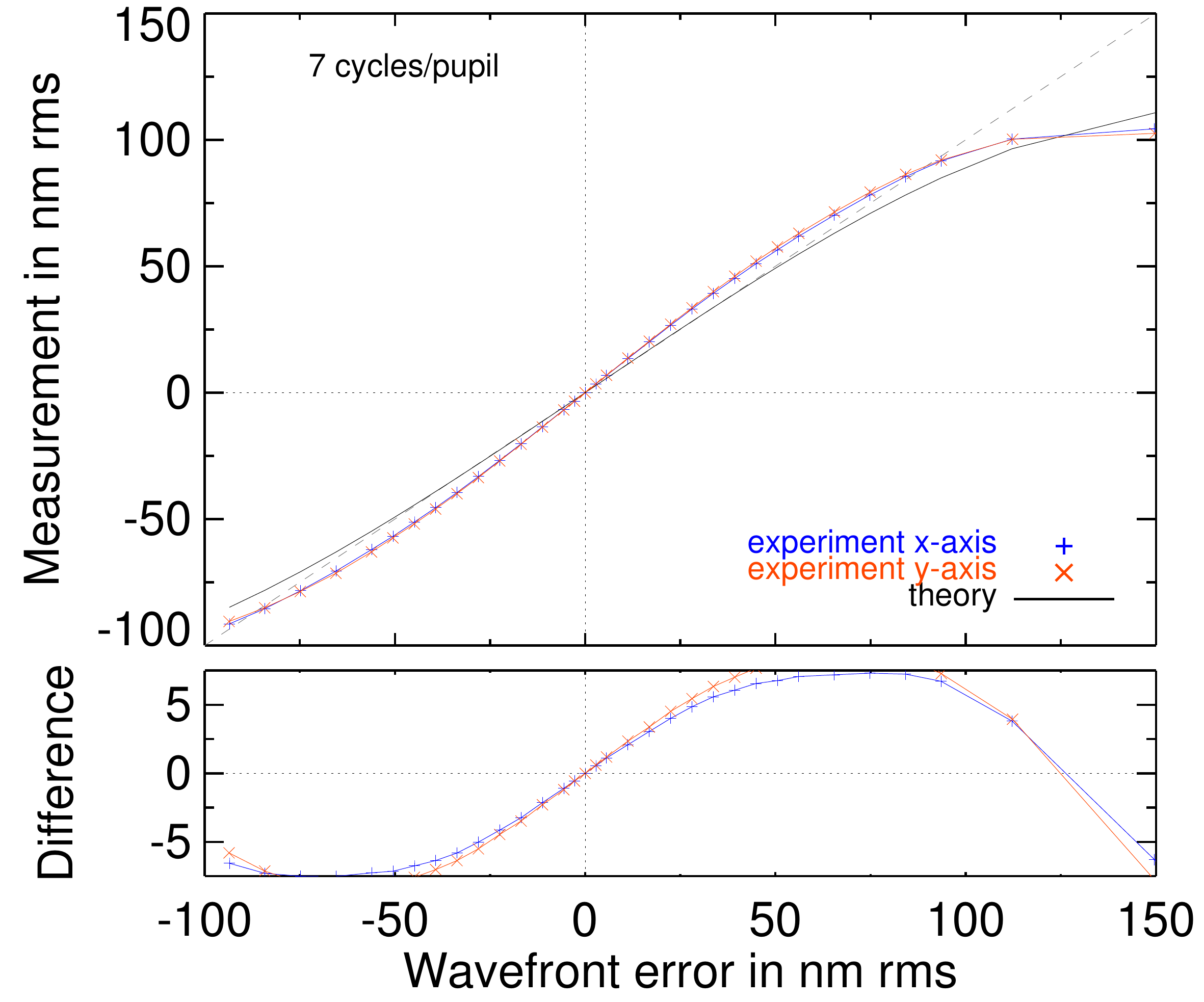}
}
\resizebox{\hsize}{!}{
\includegraphics{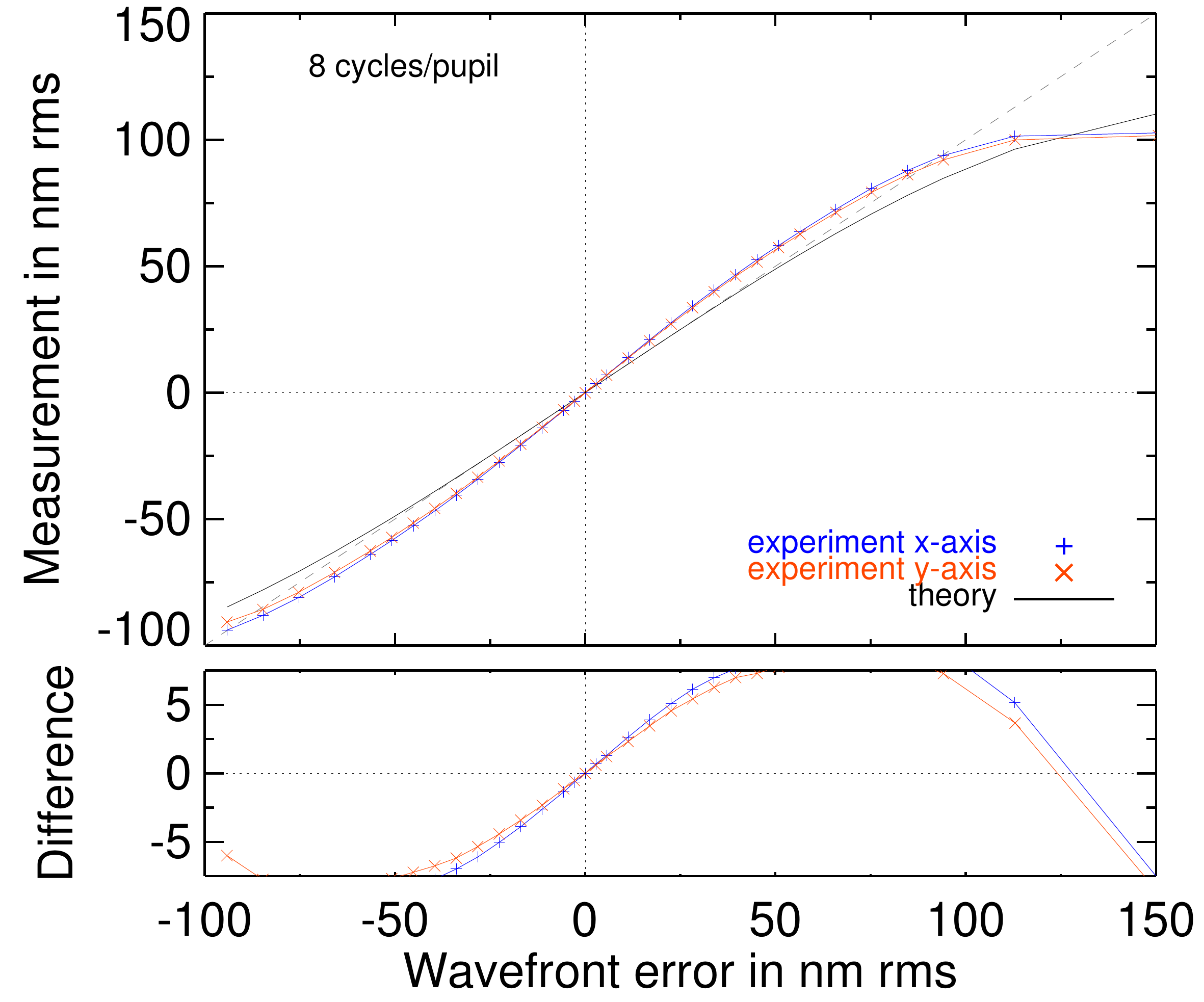}
\includegraphics{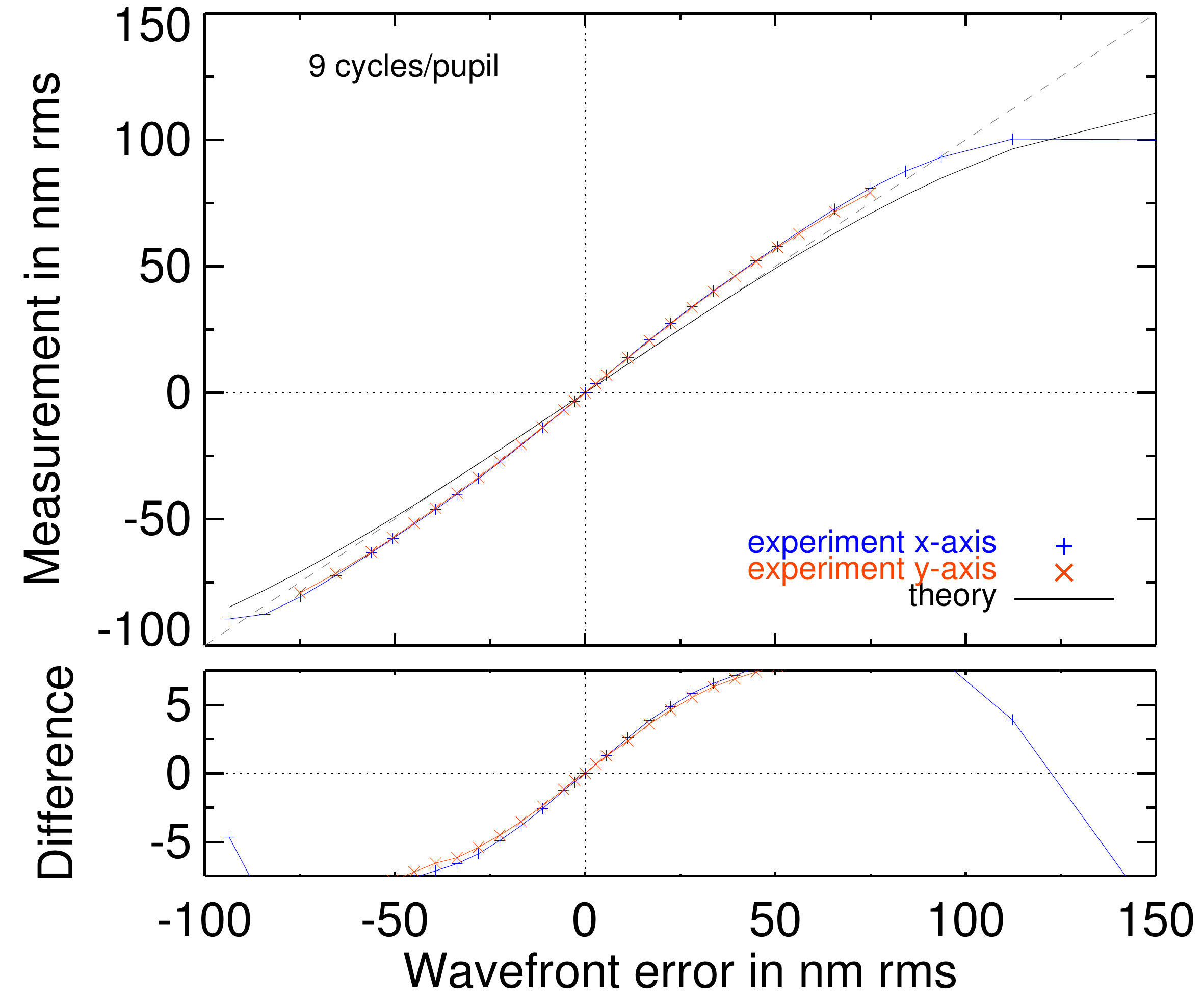}
\includegraphics{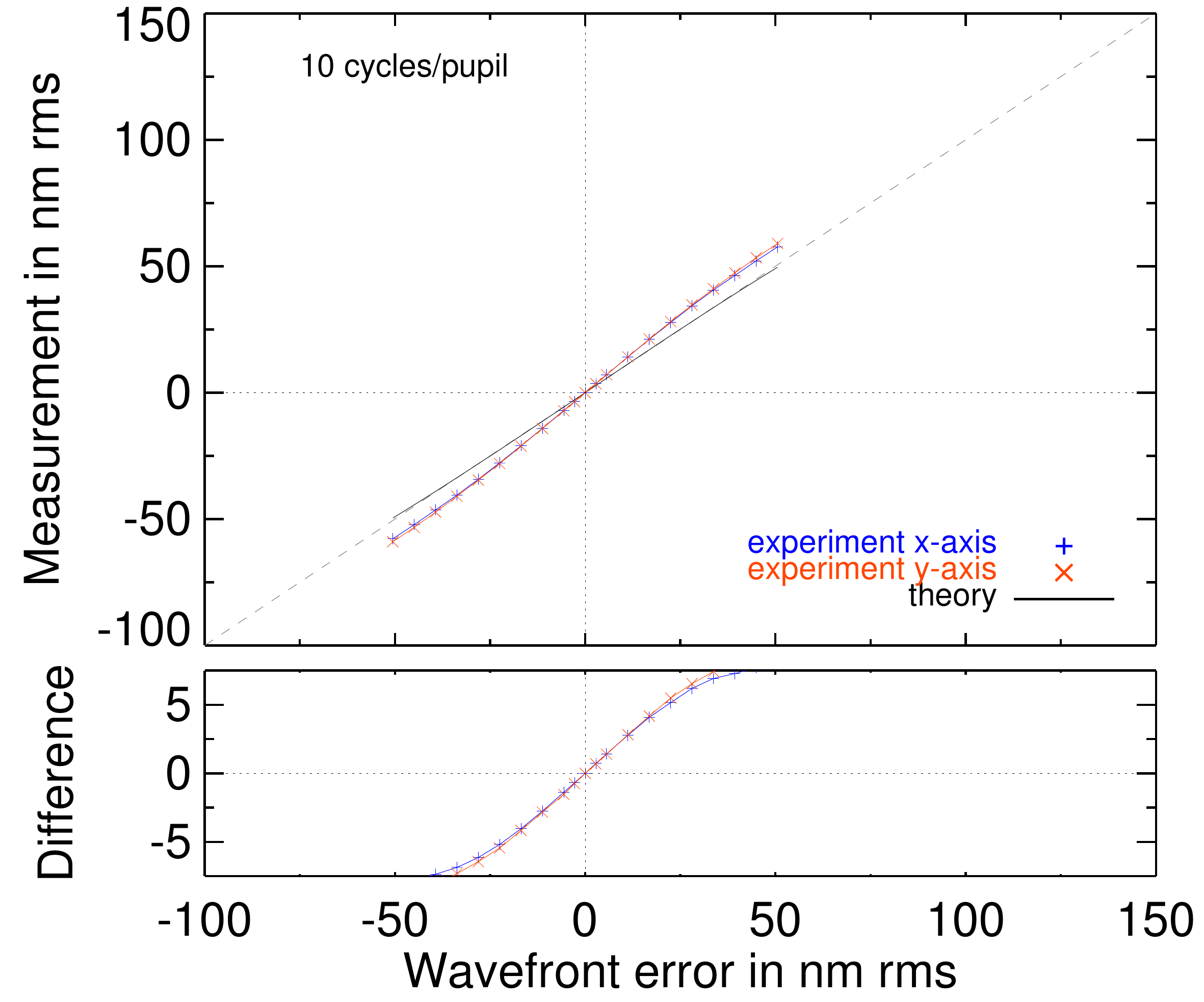}
}
\caption{\textbf{Top plot}: Theoretical and experimental aberration measurements for different Fourier modes in x- and y-axis directions as a function of the wavefront error in the entrance pupil plane. \textbf{Bottom plot}: Difference between the experiment and the theory.}
\label{fig:fourier_dir=xy_mode}
\end{figure*}

Figure \ref{fig:fourier_dir=xy_mode} presents the response curves for ZELDA with Fourier modes, showing again a very good match between the theoretical and experimental values. As for the Zernike modes, the discrepancies between theory and experiment for the Fourier modes are believed to be related to the errors in the calibration of the sensitivity factor. Our concept allows accurately measuring the aberrations corresponding to the bright speckles at the smallest separations from the star position in the coronagraphic images, thus enabling the observation of the closest and faintest planetary companions around the host star down to the raw contrast limit set by the coronagraph. 

\bibliographystyle{aa}
\bibliography{zelda2}

\end{document}